\documentclass[prd,twocolumn,superscriptaddress,floatfix,nofootinbib]{revtex4-2}
\pdfoutput=1 % if your are submitting a pdflatex (i.e. if you have
\usepackage[T1]{fontenc} % if needed
\usepackage{amsmath,amssymb,amsfonts,amsthm}
\usepackage{graphicx}
\usepackage[usenames,dvipsnames]{color}
\usepackage{enumitem}
\usepackage{verbatim}
\usepackage[percent]{overpic}
\usepackage{rotating}
\usepackage{hyperref}
\usepackage{array}
\usepackage{mathtools}
\usepackage{lmodern}
\usepackage[normalem]{ulem}
\usepackage{braket}
\usepackage{tensor}
\usepackage{dsfont}
\usepackage{bm}
\usepackage{tikz}

\usepackage{mathrsfs}

\usetikzlibrary{decorations.pathmorphing}
\usepackage{CJKutf8}

\newcommand{\tcr}{\textcolor{red}}

\usetikzlibrary{positioning}
\usetikzlibrary{calc,through,backgrounds}

\theoremstyle{definition}
\newtheorem{definition}{Definition}[section]
\newtheorem{theorem}{Theorem}[section]

\newcommand{\kako}[1]{\left( #1 \right)}
\newcommand{\kagikako}[1]{\left[ #1 \right]}
\newcommand{\ts}[1]{ _{\text{#1}} }
\newcommand{\Bigkako}[1]{\Big( #1 \Big)}

\newcommand{\Bigkagikako}[1]{\Big[ #1 \Big]}

\newcommand{\erf}{\text{erf}}
\newcommand{\erfc}{\text{erfc}}
\newcommand{\erfi}{\text{erfi}}

\allowdisplaybreaks

\DeclareMathOperator{\Tr}{Tr}

\newcommand{\dd}{\text{d}}
\newcommand{\bk}{{\bm{k}}}
\newcommand{\bx}{{\bm{x}}}

\newcommand{\id}{\mathds{1}}
\newcommand{\sx}{\mathsf{x}}
\newcommand{\sy}{\mathsf{y}}

\newcommand{\ii}{\mathsf{i}}

\newcommand{\kk}{|\bm{k}|}
\newcommand{\AAA}{\text{A}}
\newcommand{\BB}{\text{B}}

\begin{document}

\title{Extraction of entanglement from quantum fields with entangled particle detectors}

\author{Dyuman Bhattacharya}
\email{d7bhatta@uwaterloo.ca}
\affiliation{Department of Physics and Astronomy, University of Waterloo, Waterloo, Ontario, N2L 3G1, Canada}

\author{Kensuke Gallock-Yoshimura}
\email{kgallock@uwaterloo.ca} 

\affiliation{Department of Physics and Astronomy, University of Waterloo, Waterloo, Ontario, N2L 3G1, Canada}

\author{Laura J. Henderson}
\affiliation{Centre for Engineered Quantum Systems, School of Mathematics and Physics,
The University of Queensland, St. Lucia, Queensland 4072, Australia}
\email[]{laura.henderson@uq.edu.au}
\affiliation{Department of Physics and Astronomy, University of Waterloo, Waterloo, Ontario, N2L 3G1, Canada}

%\author{Erickson Tjoa}
%\email{e2tjoa@uwaterloo.ca}
%\affiliation{Department of Physics and Astronomy, University of Waterloo, Waterloo, Ontario, N2L 3G1, Canada}
%\affiliation{Institute for Quantum Computing, University of Waterloo, Waterloo, Ontario, N2L 3G1, Canada}

\author{Robert B. Mann}
\email{rbmann@uwaterloo.ca}
\affiliation{Department of Physics and Astronomy, University of Waterloo, Waterloo, Ontario, N2L 3G1, Canada}
%\affiliation{Perimeter Institute for Theoretical Physics,  Waterloo, Ontario, N2L 2Y5, Canada}

\begin{abstract}
We consider two initially entangled Unruh-DeWitt particle detectors and examine how the initial entanglement changes after interacting with a quantum scalar field. 
As initially nonentangled detectors extract entanglement from the field, entangled detectors also can gain more entanglement so long as they are weakly correlated at the beginning. 
For initially sufficiently entangled detectors, only degradation takes place. 
We then apply our analysis to a gravitational shockwave spacetime and show that a shockwave could enhance the initial entanglement of weakly entangled detectors.  Moreover, we find that this enhancement can occur for greater detector separations than in Minkowski spacetime. 
\end{abstract}

\maketitle
\flushbottom

\section{Introduction}

It is well-known by now that a vacuum state of a quantum field is an entangled state \cite{summers1985bell, summers1987bell}. 
One can examine such entanglement by making use of a particle detector, which is a nonrelativistic first-quantized system locally coupled to the field. 
A commonly used model   is a two-level quantum system known as an  Unruh-DeWitt (UDW) detector \cite{Unruh1979evaporation, DeWitt1979}, which is known to capture the essence of light-matter interactions when no angular momentum is exchanged \cite{EMM.wavepacket, Alhambra.CasimirForces,pozas2016entanglement}. 
For example, two initially uncorrelated UDW detectors (e.g., both in their ground states) can become  entangled after interacting with a quantum scalar field even when they are causally disconnected. 
In other words, the detectors extract entanglement from the field without directly exchanging quanta. 
Such a protocol is known as \textit{entanglement harvesting} \cite{Valentini1991nonlocalcorr, reznik2003entanglement, reznik2005violating, Steeg2009, pozas2015harvesting, Maeso.state.covariance}, and the amount of harvested entanglement is sensitive to the state of motion of the detectors and the geometry of underlying spacetime \cite{smith2016topology, kukita2017harvesting,henderson2018harvesting,ng2018AdS,henderson2019entangling,cong2020horizon,FinnShockwave, Liu.acceleration, Diki.inertial}. 

One can also think of initially entangled detectors and ask how much their initial entanglement changes as a function of their motion and the structure of spacetime. 
If their initial entanglement   is maximal  (e.g., a Bell state) then their entanglement will necessarily 
 decrease since after the interaction the detectors will be correlated with the field, and the monogamy of entanglement tells us that the detectors must lose their initial correlation. 
In general, the phenomenon of losing initial entanglement is known as \textit{entanglement degradation}. 
This phenomenon has been known for quite some time for   accelerating observers~\cite{Alsing.teleportation, FuentesAliceFalls, AlsingDiracFields}, with a number of subsequent studies~\cite{Lin.Disentanglement2008, Lin.Temporal.2009, Landulfo2009suddendeath, Doukas.orbit.PhysRevA.81.062320, Landulfo.SuddenChange, Ostapchuk.entanglement.dynamics, EMM.Jorma.Firewall, Rodriguez.finite.time, GallockEntangledDetectors, Chowdhury.FateEntanglement, Soares.Entanglement.dynamics} carried out for entangled UDW-type detectors. 

Entanglement harvesting and degradation suggest that two initially entangled detectors can either gain or lose entanglement after interacting with the field. 
One then can ask what  the conditions are for enhancement and degradation of entanglement. 
Such a question was   considered    for two inertial harmonic-oscillator type UDW detectors in Minkowski spacetime~\cite{Lin.Temporal.2009}, in which
 the time dependence of  the initial entanglement was analyzed by using the quantum Langevin equation with an assumption that the detectors interact with the field for an  infinitely long time after being suddenly switched on. 
The initial entanglement was found to decrease with time,  vanishing at the end. 

Here we instead consider how much the initial entanglement of a pair of UDW detectors changes after interacting with a field for a finite duration of time, working perturbatively in the field coupling. 
It is known that such a method has a technical difficulty: in a perturbative analysis, the density matrix of entangled pointlike UDW detectors contains  divergent elements \cite{EMM.Jorma.Firewall}. 
One way to circumvent this issue is to use finite-sized detectors. 
Although one must face tedious calculations in a generic curved spacetime, inertial detectors in Minkowski spacetime are tractable and yet the fate of entanglement in such a simple setting has not been examined. 
Another route is to apply an approximation to a measure of entanglement \cite{Chowdhury.FateEntanglement}: the divergent elements in the density matrix only appear in the higher order terms in the coupling constant.
Quantifying   the amount of entanglement between the detectors via  negativity, we can ignore such a divergent term for sufficiently small coupling. 
However, this approximation is valid only for initially sufficiently entangled detectors, and cannot be applied to initially weakly entangled ones. 

We investigate in this paper the extraction of vacuum entanglement by two entangled inertial detectors in Minkowski spacetime, considering situations in which the detectors are initially nearly maximally entangled and initially weakly entangled, complementing previous work in this subject~\cite{pozas2015harvesting} for initially separable detectors. 
Introducing finite-sized UDW detectors to make their density matrix well-defined, we employ  concurrence as a measure of entanglement. We examine the effects of the divergent terms on this quantity, and provide an approximation that consistently allows us to neglect such terms. 
%therefore analyze the behavior of entanglement between two inertial By introducing finite-sized UDW detectors to make their density matrix well-defined, we adopt an entanglement measure called concurrence to quantify the change in entanglement and argue that it needs the higher order terms in a density matrix of detectors to properly quantify entanglement.  By assuming small coupling, we can perform an approximation to the concurrence. Such an operation consequently omits the higher order terms and the aforementioned divergent element in the concurrence. A similar approximation can be performed to initially weakly entangled scenario. 
We show that these approximations are valid even for pointlike detectors, and utilize them to show that \textit{weakly entangled detectors can extract more entanglement from the field even when they are causally disconnected}. 
%As in the entanglement harvesting scenario, entangled detectors can extract more entanglement from the field even when they are causally disconnected, albeit one needs to tune detectors' energy gap and initial state. 

We then extend our analysis  to consider a gravitational shockwave spacetime~\cite{FinnShockwave} and explore the effect of a shockwave on the initial entanglement between pointlike detectors. 
We find that a shockwave can  enhance entanglement compared to Minkowski spacetime case:  the detectors are able to gain entanglement with smaller energy gaps and larger amounts of initial entanglement.

Our paper is organized as follows. 
We begin with providing a density matrix of initially entangled detectors after interacting with a quantum scalar field in Sec.~\ref{subsec:density matrix}. 
The divergence in the density matrix does not show up in the concurrence after performing approximations in Sec.~\ref{subsec:entanglement measure}. 
We then examine the fate of initial entanglement in $(3+1)$-dimensional Minkowski (Sec.~\ref{sec:minkowski results}) and shockwave spacetimes (Sec.~\ref{sec:UDW detectors in presence of a shockwave}), followed by Conclusion in Sec.~\ref{sec: conclusion}. 
We will use natural units, $\hbar = c=1$, and denote spacetime points by $\sx=(t,\bx)$.

\section{Initially entangled detectors}
\subsection{Density matrix}\label{subsec:density matrix}

We shall consider two UDW detectors A and B, linearly coupled to a quantum scalar field. 
Let $\tau_j$ ($j\in \{ \AAA, \BB \}$) be a proper time of detector-$j$ and $t$ a common time (i.e., a time parameter that describes both detectors). 
The interaction Hamiltonian in the interaction picture is given by 
\begin{align}
    \hat H\ts{I}(t)
    &=
        \dfrac{\dd \tau\ts{A}}{\dd t} 
        \hat H\ts{A}
        \big(
            \tau\ts{A}(t)
        \big)
        +
        \dfrac{\dd \tau\ts{B}}{\dd t} 
        \hat H\ts{B}
        \big(
            \tau\ts{B}(t)
        \big)\,,
\end{align}
where $\hat H_j(\tau_j)$ are 
\begin{align}
    \hat H_j(\tau_j)
    &=
        \lambda_j \chi_j(\tau_j) 
        \hat \mu_j(\tau_j) 
        \otimes 
        \int \dd^3 x\,F_j(\bx-\bx_{j})
        \hat \phi(\sx_j(\tau_j))\,.
\end{align}
Here, $\lambda_j$ is a coupling constant between detector-$j$ and the field $\hat \phi$, $\chi_j(\tau_j)$ is a switching function that specifies how a detector interacts, and $\hat \mu_j(\tau_j)$ is a monopole moment given by
\begin{align}
    \hat \mu_j(\tau_j)
    &\coloneqq
        e^{ \ii \Omega_j \tau_j } \ket{e_j}\bra{g_j}
        +
        e^{ -\ii \Omega_j \tau_j } \ket{g_j}\bra{e_j}
\end{align}
with $\Omega_j$ being an energy gap between the ground $\ket{g_j}$ and excited $\ket{e_j}$ states. 
$F_j(\bx-\bx_{j})$ is the so-called smearing function, which specifies the spatial profile and the center of mass position $\bx_{j}$ of detector-$j$. 
As an example, one can choose the Dirac delta distribution to make the detector pointlike. 

The time-evolution operator, $\hat U\ts{I}$, can be written by using a time-ordering symbol, $\mathcal{T}_t$, with respect to the common time $t$: 
\begin{align}
    \hat U\ts{I}
    &=
        \mathcal{T}_t \exp 
        \kako{
            -\ii \int_{\mathbb{R}} \dd t\,
            \hat H\ts{I}(t)
        } \\
    &=
        \hat U\ts{I}^{(0)} + \hat U\ts{I}^{(1)} + \hat U\ts{I}^{(2)} + \mathcal{O}(\lambda^3)\,,
\end{align}
where $\hat U\ts{I}^{(j)}$ is the $j$-th power of the coupling constant $\lambda$:
\begin{align}
    \hat U\ts{I}^{(0)}
    &\coloneqq \id \,, \\
    \hat U\ts{I}^{(1)}
    &\coloneqq 
        -\ii \int_{\mathbb{R}} \dd t\,\hat H\ts{I}(t)\,,\\
    \hat U\ts{I}^{(2)}
    &\coloneqq
        - \int_{\mathbb{R}} \dd t_1
        \int_{-\infty}^{t_1}\dd t_2\,
        \mathcal{T}_t [ \hat H\ts{I}(t_1) \hat H\ts{I}(t_2) ]\,.
\end{align}
 
Let us assume that the field is in the vacuum state, $\ket{0}$, and the detectors A and B are initially entangled
\begin{align}
    \ket{\Psi_{0}}
    &=
        \big(
            \alpha\ket{g\ts{A}}\ket{g\ts{B}}
            +\beta e^{\ii \theta}
            \ket{e\ts{A}} \ket{e\ts{B}}
        \big)\ket{0}\,,\label{eq:initial state}
\end{align}
where $\alpha, \beta \in [0,1]$ with $\alpha^2 + \beta^2=1$ and $\theta \in [0, 2\pi)$ is the relative phase. 
For example, $\alpha =0\,,1$ correspond to separable states while $\alpha=1/\sqrt{2}$ gives a Bell state. 
We will refer to $\alpha=1$ as the entanglement harvesting scenario. 
Unless otherwise stated, we let $\beta=\sqrt{ 1-\alpha^2 }$ throughout this paper. 

One can obtain a state of the detectors, $\rho\ts{AB}$, after the interaction by tracing out the field degree of freedom. 
Writing $\rho_0\coloneqq \ket{\Psi_0} \bra{\Psi_0}$, we obtain
\begin{align}
    \rho\ts{AB}
    &=
        \Tr_\phi [ \hat U\ts{I} \rho_0 \hat U\ts{I}^\dag ] \\
    &=
        \Tr_\phi[ \rho_0 ]
        + 
        \Tr_\phi[ \hat U\ts{I}^{(1)} \rho_0 \hat U\ts{I}^{(1)\dagger} ] \notag \\
        &\quad
        + \Tr_\phi[ \hat U\ts{I}^{(2)} \rho_0 ]
        + \Tr_\phi[ \rho_0 \hat U^{(2)\dagger} ] 
        + \mathcal{O}(\lambda^4)\,. 
\end{align}
\begin{widetext}
In the basis $\{ \ket{g\ts{A} g\ts{B}}, \ket{g\ts{A} e\ts{B} }, \ket{e\ts{A} g\ts{B} }, \ket{e\ts{A} e\ts{B} } \}$, the density matrix $\rho\ts{AB}$ can be obtained as follows \cite{EMM.Jorma.Firewall}. 
\begin{subequations}
\begin{align}
    \rho\ts{AB}
    &=
        \left[
        \begin{array}{cccc}
        r_{11} &0 &0 &r_{14}  \\
        0 &r_{22} &r_{23} &0  \\
        0 &r_{23}^* &r_{33} &0  \\
        r_{14}^* &0 &0 &r_{44}
        \end{array}
        \right]
        + \mathcal{O}(\lambda^4)\,, \label{eq:density matrix}\\
    r_{11}
    &=
        \alpha^2
        + 2 \alpha^2 \text{Re}[ J\ts{AA}^{(-+)} + J\ts{BB}^{(-+)} ]
        + 2 \alpha \sqrt{1-\alpha^2} 
        \text{Re}[ e^{\ii \theta} (J\ts{AB}^{(--)} + J\ts{BA}^{(--)}) ] 
        \label{r11}\\
    r_{14}
    &=
        \alpha \sqrt{1-\alpha^2}e^{-\ii\theta} 
        (1 + J\ts{AA}^{(-+)} + J\ts{AA}^{(+-)*} + J\ts{BB}^{(-+)} + J\ts{BB}^{(+-)*})
        + \alpha^2 ( J\ts{AB}^{(++)*} + J\ts{BA}^{(++)*} )
        + (1-\alpha^2) ( J\ts{AB}^{(--)} + J\ts{BA}^{(--)} )  \label{r14}\\
    r_{22}
    &=
        (1-\alpha^2) I\ts{AA}^{(+-)}
        +
        2 \alpha \sqrt{1-\alpha^2} \text{Re}[ e^{-\ii \theta} I\ts{AB}^{(++)} ]
        + \alpha^2 I\ts{BB}^{(-+)}  \label{r22}\\
    r_{23}
    &=
        \alpha \sqrt{1-\alpha^2} e^{\ii \theta} I\ts{AA}^{(--)} 
        + (1-\alpha^2) I\ts{BA}^{(+-)}
        + \alpha^2 I\ts{AB}^{(-+)}
        + \alpha \sqrt{1-\alpha^2} e^{-\ii\theta} I\ts{BB}^{(++)}\\
    r_{33}
    &=
        \alpha^2 I\ts{AA}^{(-+)} 
        +
        2 \alpha \sqrt{1-\alpha^2} \text{Re}[ e^{\ii \theta} I\ts{AB}^{(--)} ]
        + (1-\alpha^2) I\ts{BB}^{(+-)}  \label{r33}\\
    r_{44}
    &=
        (1-\alpha^2) 
        + 2 (1-\alpha^2) \text{Re}[ J\ts{AA}^{(+-)} + J\ts{BB}^{(+-)} ]
        + 2 \alpha \sqrt{1-\alpha^2} 
        \text{Re}[ e^{-\ii \theta} (J\ts{AB}^{(++)} + J\ts{BA}^{(++)}) ]  \label{r44}
\end{align}
\end{subequations}
where 
\begin{align}
    I_{jk}^{(pq)}
    &\coloneqq
        \lambda_j\lambda_k
        \int_{\mathbb{R}} \dd \tau_j 
        \int_{\mathbb{R}} \dd \tau_k\,
        \chi_j ( \tau_j) \chi_k( \tau_k )
        e^{ \ii (p \Omega_j \tau_j + q \Omega_k \tau_k ) }
        \mathcal{W}\big( \sx_{j}(\tau_j), \sy_{k}(\tau_k) \big)\,,  \label{Ijk}\\
    J_{jk}^{(pq)}
    &\coloneqq
        -\lambda_j \lambda_k
        \int_{\mathbb{R}} \dd \tau_j \int_{-\infty}^{t_1(\tau_j)} \dd \tau_k\,
        \chi_{j} ( \tau_j ) \chi_k ( \tau_k )
        e^{ \ii ( p\Omega_j \tau_{j} + q \Omega_k \tau_{k} ) }
        \mathcal{W}\big( \sx_{j}(\tau_j), \sy_{k}(\tau_k) \big) \label{Jjk} \,,
\end{align}
and we have used the fact that $I\ts{AB}^{(\pm \pm)}=I\ts{BA}^{(\mp \mp)*} \in \mathbb{C}$ in $r_{22}$ and $r_{33}$ respectively. 
The elements $I\ts{AB}^{(\pm \pm)}$ and $J_{kk}^{(\pm \mp)}$ appear in the density matrix \eqref{eq:density matrix} only for initially entangled detectors. 
\end{widetext}
The quantity
\begin{align}
    \mathcal{W}(\sx_j, \sy_k)
    &\coloneqq
        \int \dd^3 x
        \int \dd^3 y\,
        F_j(\bx-\bx_j)
        F_k(\bm{y}-\bm{y}_j)
        W(\sx, \sy),
\end{align}
where $W(\sx, \sy)\coloneqq \bra{0} \hat \phi(\sx) \hat \phi(\sy) \ket{0}$ is the Wightman function (two-point vacuum correlation function). 
In the above equations, the Wightman functions, $W(\sx_j(\tau_j), \sy_k(\tau_k))$, are pulled back along the trajectories of detectors-$j$ and $k$. 

%\tcb{One can show that $I\ts{AB}^{(\pm \pm)}=I\ts{BA}^{(\mp \mp)*} \in \mathbb{C}$ in $r_{22}$ and $r_{33}$. This simplifies the second and third terms in $r_{22}$ and $r_{33}$ to $2 \alpha \sqrt{1-\alpha^2} \text{Re}[ e^{-\ii \theta} I\ts{AB}^{(++)} ]$ and $2 \alpha \sqrt{1-\alpha^2} \text{Re}[ e^{\ii \theta} I\ts{AB}^{(--)} ]$, respectively. These quantities could be positive or negative.}

\subsection{Entanglement measure}\label{subsec:entanglement measure}

To quantify the entanglement between the detectors we use the \textit{concurrence}, 
$\mathcal{C}\ts{AB}$.
For density matrices of the form \eqref{eq:density matrix}, it is known that the qubits are entangled if and only if one of the following is satisfied \cite{smith2016topology}. 
\begin{align}
    |r_{14}|^2 > r_{22} r_{33}\,,
    \qquad |r_{23}|^2 > r_{11}r_{44}\,.
\end{align}
We find that $|r_{23}|^2 > r_{11}r_{44}$ can never be realized;  thereby the concurrence is defined as  
\begin{align}
    \mathcal{C}\ts{AB}
    \coloneqq 
        2 \max \{ 0,\,|r_{14}|-\sqrt{r_{22} r_{33} } \}\,.\label{eq:concurrence}
\end{align}
Note that $0 \leq \mathcal{C}\ts{AB} \leq 1$ and $\mathcal{C}\ts{AB}=0$ if and only if two detectors are not entangled. 
One can easily verify that the initial entanglement in \eqref{eq:initial state} is 
\begin{align}
    \mathcal{C}\ts{AB,0}
    \coloneqq
        2 \alpha \sqrt{1-\alpha^2} ~
        (= 2\beta \sqrt{1-\beta^2})\,.
\end{align}

Although evaluating the concurrence \eqref{eq:concurrence} may seem straightforward, inspection of \eqref{r14},  \eqref{r22}, and \eqref{r33} indicates that care must be taken in perturbatively approximating $\mathcal{C}\ts{AB}$. 
This is because each of $r_{22}$ and $r_{33}$ are of order $\lambda^2$ (and so their geometric mean  {$\sqrt{ r_{22} r_{33} }$} is of order $\lambda^2$)
whereas a perturbative expansion of $r_{14}$ 
\begin{align}
    r_{14}
    &=
        r_{14}^{(0)} + \lambda^2 r_{14}^{(2)} + \lambda^4 r_{14}^{(4)} + \mathcal{O}(\lambda^6)\,,\label{eq:r14 form}
\end{align}
yields
\begin{align}
    |r_{14}|^2
    &=
        |r_{14}^{(0)}|^2
        + \lambda^2 2 \text{Re}[ r_{14}^{(0)} r_{14}^{(2)*} ] \notag \\
        &\quad
        + \lambda^4 (|r_{14}^{(2)}|^2 + 2 \text{Re}[ r_{14}^{(0)} r_{14}^{(4)*} ] )
        + \mathcal{O}(\lambda^6)\,.\label{eq:corrected r14}
\end{align}
%\sout{and so $r_{14}^{(4)}$ cannot be neglected unless $r_{14}^{(0)}$ is sufficiently small.} \ljh{This needs a bit of clarification.  Since we're considering $|r_{14}|$ not $|r_{14}|^2$, the $\lambda^4$ term will never matter \textit{unless} $r_{14}^{(0)}$ is small, specifically $r_{14}^{(0)}<\lambda^4$ (which is exactly the case that allows us to recover regular harvesting). To clarify: ${\sqrt{a_0+a_2\lambda^2+a_4\lambda^4} \approx \sqrt{a_0}+\lambda^2a_2/(2\sqrt{a_0})}$ but ${\sqrt{a_4\lambda^4+a_6\lambda^6} \approx \lambda^2\sqrt{a_4}}$. (In the second case, only $|r_{14}^{(2)}|^2$ will matter, since technically $\lambda^4\text{Re}[r_{14}^{(0)}{r_{14}^{(4)}}^*]$ will go like $\lambda^8$ or smaller.)}
For initially separable states, (the standard harvesting protocol) $r_{14}^{(0)} = 0$, and so $|r_{14}|^2=\lambda^4|r_{14}^{(2)}|^2 + \mathcal{O}(\lambda^6)$, indicating that there values of $\alpha$, close to 0 and 1, for which $\lambda^4$ terms must make a significant contribution to $|r_{14}|$.  In order to keep the perturbative expansion of $|r_{14}|$ consistent, we will consider the cases where the detectors are nearly separable and the case where the detectors are sufficiently entangled separately.  
%\sout{but more generally we wish to understand the harvesting of entanglement for $r_{14}^{(0)} \neq 0$.} 

Another issue has to do with the structure of $r_{14}$, which from \eqref{r14} can be written as
 \begin{align}
    r_{14}^{(0)}
    &=
        \alpha
        \sqrt{1-\alpha^2} e^{-\ii\theta}\,, \label{eq:simplified r14(0)} \\
    \lambda^2 r_{14}^{(2)}
    &=
        \lambda^2 
        \Bigkagikako{
            \alpha
            \sqrt{1-\alpha^2} e^{-\ii\theta} (Y\ts{A} + Y\ts{B}) \notag \\
            &\qquad
            + \alpha^2 X\ts{AB}^*(-\Omega)
            + (1-\alpha^2) X\ts{AB}(\Omega)
        }\,,\label{eq:simplified r14(2)}
\end{align}
where 
\begin{subequations}
\begin{align}
    &\lambda^2 Y_{k}
    \equiv
        J_{kk}^{(-+)} + J_{kk}^{(+-)*} \,, \\
    &\lambda^2 X\ts{AB}(\Omega)
    \equiv
        J\ts{AB}^{(--)} + J\ts{BA}^{(--)}\,,\\
    &\lambda^2 X\ts{AB}^*(-\Omega)
    \equiv
        J\ts{AB}^{(++)*} + J\ts{BA}^{(++)*}\,.
\end{align}
\end{subequations}
The imaginary part of $Y_k$ is divergent and causes problems in the pointlike limit \cite{EMM.Jorma.Firewall}. 
Although  it can be regularized by introducing a smearing function, we shall demonstrate that it does not appear in the concurrence under the approximations we consider.

\subsection{Approximating the Concurrence}\label{subsec:approximation in concurrence}

We consider three scenarios: \\

\noindent
(i) Initial weak entanglement with $\alpha \approx 0$; \\
(ii) Initial weak entanglement with $\alpha \approx 1$; \\
(iii) Initial sufficient  entanglement  \\

\noindent
$\bullet$ \textit{Initial weak entanglement: $\alpha \approx 0$}\\

Let us assume that $\alpha \ll 1$ and perform a series expansion of Eq.~\eqref{eq:corrected r14} in terms of $\lambda^n \alpha^m$: 
\begin{align}
    |r_{14}|^2
    &=
        \underbrace{\alpha^2 (1-\alpha^2)}_{\text{initial}}
        + \underbrace{ 2 \lambda^2 \alpha \text{Re}[ X\ts{AB}(\Omega) e^{\ii \theta} ] }_{\text{neutral}} \notag \\
        &\quad+
        \underbrace{ \lambda^4 |X\ts{AB}(\Omega)|^2  }_{\text{harvesting}}
        + \underbrace{2 \lambda^2 \alpha^2 \text{Re}[Y\ts{A}+Y\ts{B}]}_{\text{degradation}} \notag \\
        &\quad+ \mathcal{O}(\lambda^n \alpha^m)\,.
        \qquad (n+m > 4) \label{eq:approx r14 alpha 0}
\end{align}
As one can see, $r_{14}^{(4)}$ and the divergent element $\text{Im}[Y_k]$ do not exist in the lower-order terms. \\

\noindent
$\bullet$ \textit{Initial weak entanglement: $\alpha \approx 1$}\\

In the same manner, one can obtain a similar expression for $\alpha \approx 1$. 
For simplicity, let us use $\beta=\sqrt{1-\alpha^2}$ and expand Eq.~\eqref{eq:corrected r14} around $\beta =0$:
\begin{align}
    |r_{14}|^2
    &=
        \underbrace{\beta^2 (1-\beta^2)}_{\text{initial}}
        + \underbrace{ 2 \lambda^2 \beta \text{Re}[ X^*\ts{AB}(-\Omega) e^{\ii \theta} ] }_{\text{neutral}} \notag \\
        &\quad+
        \underbrace{ \lambda^4 |X\ts{AB}(-\Omega)|^2  }_{\text{harvesting}}
        + \underbrace{2 \lambda^2 \beta^2 \text{Re}[Y\ts{A}+Y\ts{B}]}_{\text{degradation}} \notag \\
        &\quad
        + \mathcal{O}(\lambda^n \beta^m)\,.
        \qquad (n+m > 4) \label{eq:approx r14 alpha 1}
\end{align}
As before, the elements $r_{14}^{(4)}$ and $\text{Im}[Y_k]$ are absent. 
Note that $\beta=0$ reduces to the entanglement harvesting scenario: $|r_{14}|=\lambda^2 |X\ts{AB}(-\Omega)|$.

In both cases $\alpha \approx 0$ and 1, the approximated $|r_{14}|^2$ consists of four parts. 
The initial entanglement contribution is $\alpha^2 (1-\alpha^2)$ in \eqref{eq:approx r14 alpha 0} [equivalent to $\beta^2 (1-\beta^2)$ in \eqref{eq:approx r14 alpha 1}]. The third term containing $X\ts{AB}(\Omega)$ is the part contributing to entanglement harvesting, and so will  enhance entanglement. 
The last term, containing $\text{Re}[Y\ts{A} + Y\ts{B}]$, depends only on each detector and not on their correlations. 
This term corresponds to an outcome of local operations and so  never enhances entanglement. 
Indeed, one can straightforwardly show that $\lambda^2\text{Re}[Y_j] = -(I_{jj}^{(-+)}+I_{jj}^{(+-)})/2 \leq 0$ and so degrades entanglement.
The second term, which we denote as the neutral contribution, can be either positive or negative. By choosing the relative phase $\theta$ appropriately, we can   ensure that $\text{Re}[ e^{-\ii \theta} X\ts{AB} ] > 0$, which enhances entanglement.

In summary, we see from \eqref{eq:concurrence} that the detectors gain entanglement after the interaction provided  the harvesting part is greater than other two parts in $|r_{14}|$ and $\sqrt{r_{22} r_{33}}$.  
 Indeed, if the harvesting part does not exceed \textit{both} the initial and degradation contributions to  $|r_{14}|$, then entanglement degradation is inevitable, regardless of the value of $\sqrt{r_{22} r_{33}}$.
This is not the case for harvesting: even if the harvesting contribution is greater than the other two in $|r_{14}|$,    entanglement enhancement is not ensured because this depends on $\sqrt{r_{22} r_{33}}$. \\

\noindent
$\bullet$ \textit{Initial sufficient entanglement }\\

In this approximation 
we assume $|r_{14}^{(0)}|^2 \gg \mathcal{O}(\lambda^4)$ and 
 write Eq.~\eqref{eq:corrected r14} explicitly as~\cite{Chowdhury.FateEntanglement}
\begin{widetext}
\begin{subequations}
\begin{align}
    &|r_{14}^{(0)}|^2
    =
        \alpha^2(1-\alpha^2)\,,\\
    &\text{Re}[ r_{14}^{(0)} r_{14}^{(2)*} ]
    =
        \alpha \sqrt{1-\alpha^2}
        \Bigkagikako{
            \alpha \sqrt{1-\alpha^2} \text{Re}[Y\ts{A} + Y\ts{B}]
            +
            \Bigkako{
                \alpha^2 \text{Re}[ X\ts{AB}^*(-\Omega) e^{\ii \theta} ]
                +
                (1-\alpha^2) \text{Re}[ X\ts{AB}(\Omega) e^{\ii \theta} ]
            }
        }\,,\\
    &|r_{14}^{(2)}|^2
    =
        \left|
            \alpha^2 X\ts{AB}^*(-\Omega) + (1-\alpha^2) X\ts{AB}(\Omega)
        \right|^2
        + \alpha^2 (1-\alpha^2) |Y\ts{A}+Y\ts{B}|^2 \notag \\
        &\qquad
        + 2\alpha \sqrt{1-\alpha^2}
        \Bigkako{
            \alpha^2 \text{Re}[ X\ts{AB}^*(-\Omega) (Y\ts{A}+Y\ts{B})^* e^{\ii \theta} ]
            +
            (1-\alpha^2) \text{Re}[ X\ts{AB}(\Omega) (Y\ts{A}+Y\ts{B})^* e^{\ii \theta} ]
        }\,.\label{eq:explicit higher order terms in r14}
\end{align}
\end{subequations}

% \begin{align}
%     |r_{14}|^2
%     &=
%         \alpha^2(1-\alpha^2)
%         +2 \alpha \sqrt{1-\alpha^2} \lambda^2
%         \Bigkagikako{
%             \alpha \sqrt{1-\alpha^2} \text{Re}[Y\ts{A} + Y\ts{B}]
%             +
%             \Bigkako{
%                 \alpha^2 \text{Re}[ a e^{-\ii \theta} ]
%                 +
%                 (1-\alpha^2) \text{Re}[ b e^{-\ii \theta} ]
%             }
%         } \notag \\
%         &\quad + \lambda^4
%         \Bigkagikako{
%             \left|
%                 \alpha^2 a + (1-\alpha^2)b
%             \right|^2
%             + \alpha^2 (1-\alpha^2) |Y\ts{A}+Y\ts{B}|^2
%             + 2\alpha \sqrt{1-\alpha^2}
%             \Bigkako{
%                 \alpha^2 \text{Re}[ a z^* e^{-\ii \theta} ]
%                 +
%                 (1-\alpha^2) \text{Re}[ b z^* e^{-\ii \theta} ]
%             }
%         }\,.
% \end{align}
\end{widetext}
Omitting the $\lambda^4$ terms, we obtain 
\begin{align}
    |r_{14}|^2
    &\approx
        |r_{14}^{(0)}|^2
        + \lambda^2 2 \text{Re}[ r_{14}^{(0)} r_{14}^{(2)*} ]\label{eq:approx r14 sufficient}
\end{align}
which is also independent of $r_{14}^{(4)}$ and $\text{Im}[Y_k]$. 
Obviously this approximation does not work for $\alpha =0$ and 1 since it cannot recover the entanglement harvesting results. 
In fact, as in the previous approximations, Eq.~\eqref{eq:approx r14 sufficient} consists of three components: the initial entanglement, the harvesting contribution, and the degradation contribution. However now 
the harvesting contribution does not contain $|X\ts{AB}(\Omega)|^2$ or $|X\ts{AB}^*(-\Omega)|^2$. 
Note that for $\alpha \ll 1$ the leading
term in \eqref{eq:explicit higher order terms in r14} yields the harvesting
contribution in~\eqref{eq:approx r14 alpha 0} (and likewise for $\beta \ll 1$ the harvesting contribution in \eqref{eq:approx r14 alpha 1} is recovered).

%In summary, one has to pay a great attention to $|r_{14}|$ in the concurrence to avoid pathological results. 

Henceforth we shall consider these  approximations (which are applicable even to pointlike detectors) in evaluating the concurrence in the scenarios we consider.

\subsection{Smeared detectors in a flat spacetime}\label{subsubsec: Smeared detectors}

We now compute the elements in the density matrix \eqref{eq:density matrix} in $(3+1)$-dimensional Minkowski spacetime. 
We assume that the detectors are identical (have the same shape and internal structure) and are both at rest in a single frame of reference. 
We choose the switching function, $\chi_j(\tau_j)$, and smearing function, $F_j(\bx-\bx_j)$, to be Gaussian functions: 
\begin{align}
    &\chi_j(\tau_j)
    =
        e^{ -(\tau_j - \tau_{j,0})^2/T^2 }
        \,, \\
    &F_j(\bx - \bx_j)
    =
        \dfrac{1}{ ( \sqrt{\pi} \sigma )^3 }
        e^{ -(\bx - \bx_j )^2/\sigma^2 }\,, \label{eq:smearing function}
\end{align}
where  $\tau_{j,0}$ and $\bx_{j}$ are their respective centers  
and $T$ and $\sigma$ are
their respective temporal and spatial widths.
 
\begin{figure*}[t]
    \centering
    \includegraphics[width=\textwidth]{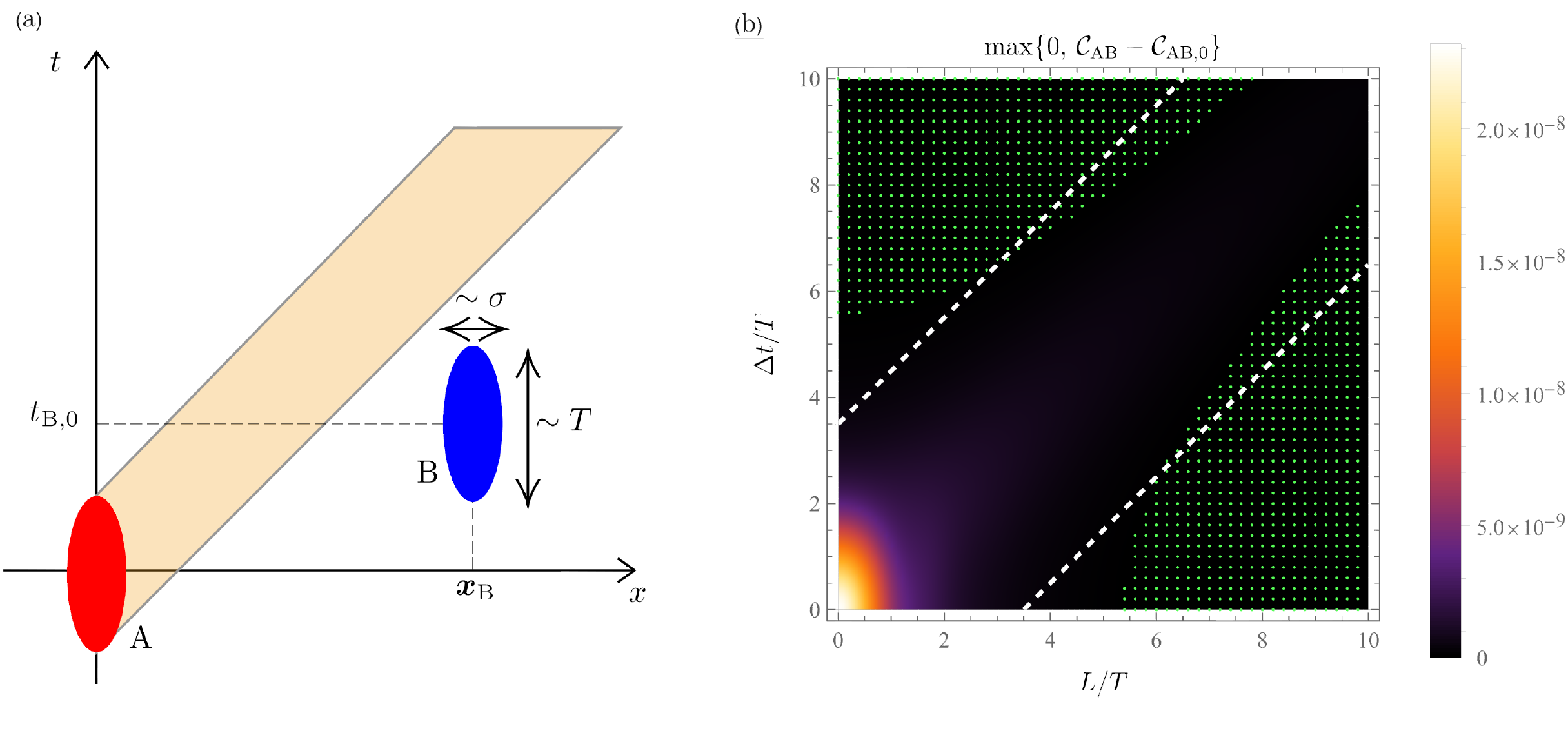}\\
    \caption{
    (a) Finite-sized detectors in Minkowski spacetime. 
    The center of mass of detector-B is located at $(t\ts{B,0}, \bx\ts{B})$;  each detector has the respective spatial and temporal Gaussian widths $\sigma$ and $T$. 
    The diagonal orange-shaded strip represents the null trajectories from detector-A;  detector-A can directly signal to B by exchanging field quanta once the support of detector-B crosses this region. 
    (b) A plot of the positive part of difference $\mathcal{C}\ts{AB} - \mathcal{C}\ts{AB,0}$ between the concurrence before and after the interaction, with  $\beta=10^{-15}, \lambda=1/10, \Omega T=5, \sigma/T=0$, and $\theta=0$. 
    The green dots represent $\mathcal{C}\ts{AB}-\mathcal{C}\ts{AB,0}\leq 0$, which means the initial entanglement is degraded, and the dashed lines indicate the null trajectory corresponding to the one in (a).}
    \label{fig:spacetime diagram and plot}
\end{figure*}

As we show in the  Appendix, the quantities  $J_{kk}^{(-+)} + J_{kk}^{(+-)*}, J\ts{AB}^{(++)*}$, and $J\ts{AB}^{(--)}$ in $r_{14}$, and $I_{jk}^{(\pm \pm)}$ in $r_{22}$ and $r_{33}$ are then given by 
\begin{widetext}
\begin{subequations}
\begin{align}
    &\lambda_k^2 Y_k
    \equiv
    J_{kk}^{(-+)} + J_{kk}^{(+-)*}  \notag \\
    &\qquad=
        -\dfrac{ \lambda_k^2 e^{ -T^2 \Omega^2/2 } }{ 8\pi (1 + \sigma^2/T^2)^{3/2} }
        \kagikako{
            2 \sqrt{1 + \sigma^2/T^2}
            +
            e^{ T^2 \Omega^2/2(1 + \sigma^2/T^2) }
            \sqrt{2\pi} T \Omega 
            \erf 
            \kako{
                \dfrac{ T \Omega }{ \sqrt{ 2(1 + \sigma^2/T^2) } }
            }
        } \notag \\
        &\qquad
        - \ii 
        \dfrac{ \lambda_k^2 T^2 }{8\pi }
        \int_0^\infty \dd \kk\,
        \kk 
        e^{ -\kk^2 \sigma^2/2 }
        \kagikako{
            e^{ -T^2(\kk-\Omega)^2/2 } 
            \erfi 
            \kako{
                \dfrac{ T(\kk - \Omega) }{\sqrt{2} }
            }
            - 
            e^{ -T^2(\kk+\Omega)^2/2 } 
            \erfi 
            \kako{
                \dfrac{ T(\kk + \Omega) }{\sqrt{2} }
            }
        }\,, \label{eq:problematic term with smearing} \\
    &J\ts{AB}^{(--)}
    =
        -\dfrac{\lambda\ts{A}\lambda\ts{B} T^2}{8 \pi L}
        e^{ -\ii \Omega (t\ts{A,0} + t\ts{B,0}) }
        e^{ -T^2 \Omega^2/2 } 
        \int_0^\infty \dd \kk\,
        e^{ -\kk^2 (\sigma^2+T^2)/2 }
        e^{ \ii \Delta t \kk }
        \erfc 
        \kako{
            \dfrac{ \Delta t + \ii T^2 \kk }{ \sqrt{2} T }
        }
        \sin (\kk L)\,, \label{eq:JABmm}\\
    &I_{kk}^{(-+)}
    = 
        \dfrac{\lambda_k^2 e^{ -T^2 \Omega^2/2 } }{ 8\pi (1 + \sigma^2/T^2)^{3/2} } 
        \kagikako{
            2 \sqrt{ 1 + \sigma^2/T^2 }
            -
            \sqrt{ 2\pi } T \Omega 
            e^{ T^2 \Omega^2/2( 1+\sigma^2/T^2 ) }
            \erfc 
            \kako{
                \dfrac{T \Omega}{ \sqrt{ 2(1 + \sigma^2/T^2) } }
            }
        } \label{eq:Ikkpm} \,,\\
    &I\ts{AB}^{(++)}
    % =
    %     \dfrac{ \lambda\ts{A}\lambda\ts{B} T e^{ \ii \Omega ( t\ts{A}+t\ts{B}) } 
    %     e^{ -T^2 \Omega^2 /2 }  }{ 4 \pi L \sqrt{ 1 + \sigma^2/T^2 } }
    %     \sqrt{ \dfrac{\pi}{2} } 
    %     \kagikako{
    %         e^{ -\alpha_-^2 } ( \ii + \erfi(\alpha_-) )
    %         -
    %         e^{ -\alpha_+^2 } ( \ii - \erfi(\alpha_+) )
    %     }\,, \\
    =
        \dfrac{ \ii \lambda\ts{A}\lambda\ts{B} T e^{ \ii \Omega ( t\ts{A,0}+t\ts{B,0}) } 
        e^{ -T^2 \Omega^2 /2 }  }{ 8 \pi L \sqrt{ 1 + \sigma^2/T^2 } }
        \sqrt{ \dfrac{\pi}{2} } 
        \kagikako{
            e^{ -\Gamma_-^2 } \erfc(\ii \Gamma_-)
            -
            e^{ -\Gamma_+^2 } \erfc( -\ii \Gamma_+)
     \label{eq:IABpp}   }
\end{align}

\end{subequations}
\end{widetext}
where $L\coloneqq |\bm{x}\ts{B}-\bm{x}\ts{A}|$ and $\Delta t\coloneqq t\ts{B,0}-t\ts{A,0}$ are spatial and temporal separation of the detectors' Gaussian peaks [see Fig.~\ref{fig:spacetime diagram and plot}(a)], and
\begin{align}
    \Gamma_\pm
    &\coloneqq
        \dfrac{ L \pm \Delta t }{ T \sqrt{ 2( 1 + \sigma^2 /T^2 ) } }\, .
\end{align}
We remark that $J\ts{AB}^{(++)*}$ can be obtained from $J\ts{AB}^{(--)}$ by $\Omega \to -\Omega$ and then taking a complex conjugate. 
In the same manner, one finds $I\ts{BA}^{(--)}=I\ts{AB}^{(++)*}$. 
Note that the real part of \eqref{eq:problematic term with smearing} is always nonpositive, which means that $\text{Re}[Y\ts{A} + Y\ts{B}]$ in $|r_{14}|$ acts as noise that inhibits the detectors from gaining entanglement. 
In fact, in  units of $\Omega$, one can verify from \eqref{eq:problematic term with smearing} that $\text{Re}[Y_k]\sim -T\Omega$ when $T\Omega \gg 1$ for $\sigma \geq 0$. 
This suggests that the longer the interaction is, the more leakage of entanglement occurs, and thereby entanglement extraction becomes more difficult as the interaction duration increases. 

Thanks to the smearing function, all the elements in the density matrix are finite and well-defined. 
One can then perform the approximations in Sec.~\ref{subsec:entanglement measure} and take a limit $\sigma \to 0$ if pointlike detectors are chosen. 
In the following sections, we will adopt pointlike detectors. 
The qualitative behavior for smeared detectors is not so different from the pointlike ones.

\begin{figure*}[t]
    \centering
    \includegraphics[width=\textwidth]{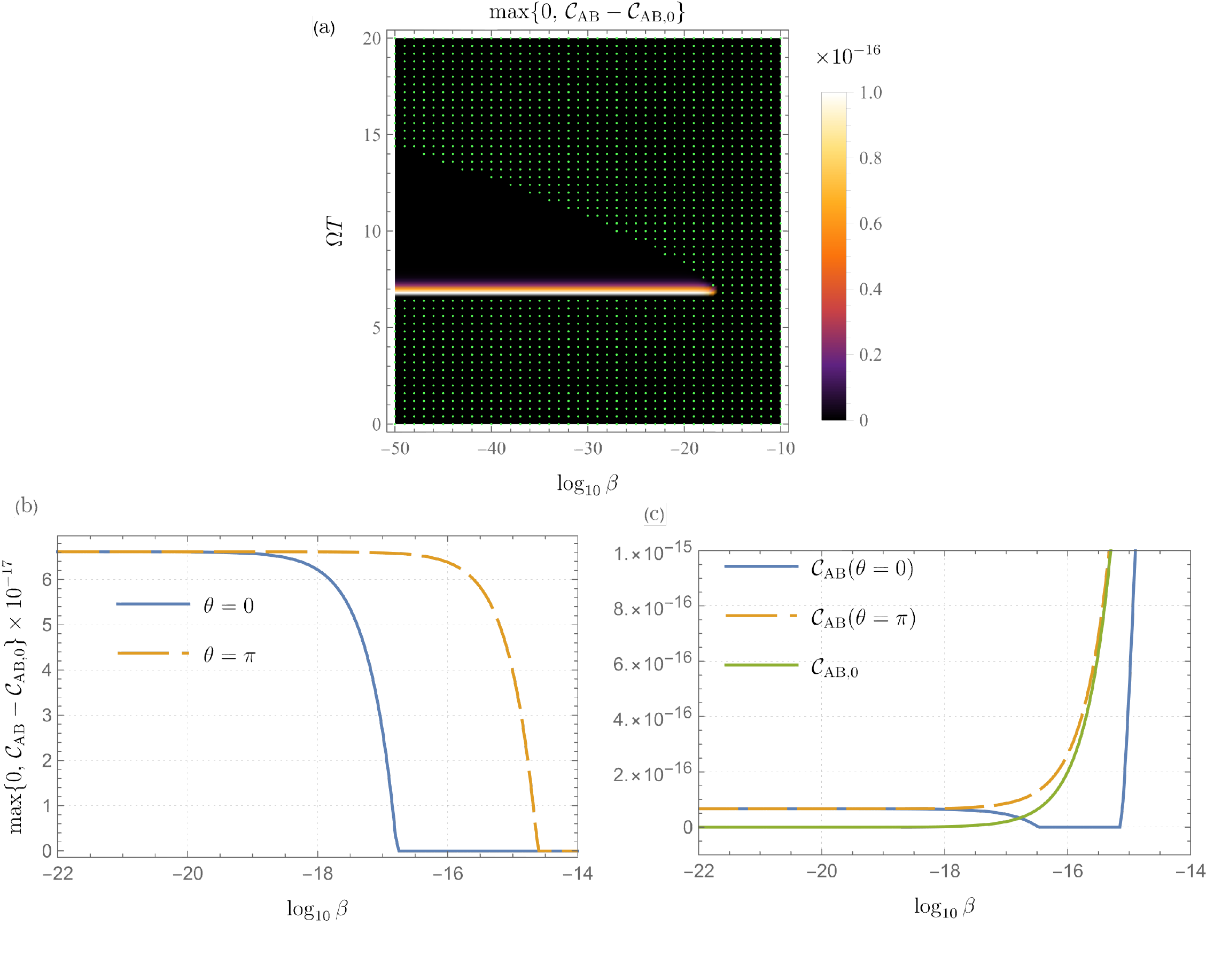}\\
    \caption{
    (a) The positive part of $\mathcal{C}\ts{AB} - \mathcal{C}\ts{AB,0}$ as a function of $\Omega T$ and $\log_{10}\beta$ when $\lambda=1/10, L/T=7, \Delta t/T=0,  \sigma/T=0$, and $\theta=0$. 
    The green dots represent degradation. 
    The detectors can gain entanglement when $\beta$ is small enough. 
    The range of $\Omega T$ that enables the detectors to gain entanglement enlarges as $\beta \to 0$. 
    (b) $\max\{ 0,\,\mathcal{C}\ts{AB}-\mathcal{C}\ts{AB,0} \}$ as a function of $\log_{10}\beta$ for $\lambda=1/10, L/T=7, \Delta t/T=0,  \sigma/T=0$, and $\Omega T=7$. 
    The blue curve shows a slice in (a) at $\Omega T=7$, whereas the orange dashed curve represents the case when $\theta=\pi$. 
    (c) Each component $\mathcal{C}\ts{AB}$ and $\mathcal{C}\ts{AB,0}$ in (b) with the same parameters. }
    \label{fig:Omega and beta plot}
\end{figure*}

\section{The fate of entanglement between inertial detectors}\label{sec:minkowski results}

In this section, we consider pointlike ($\sigma \to 0$) detectors in $(3+1)$-dimensional Minkowski spacetime and see how the initial entanglement changes after the interaction. 
% We then examine the validity of approximation in the concurrence described in Sec.~\ref{subsubsec:approximation in concurrence}. 

As shown in Fig.~\ref{fig:spacetime diagram and plot}(a), suppose detector-A's center of mass is located at the origin and the peak of Gaussian switching is at $t=0$, with detector-B's position in a spacetime, $(t\ts{B,0}, \bx\ts{B})$ fixed but variable. 
%We remind the readers that $L\coloneqq |\bx\ts{B}-\bx\ts{A}|$ and $\Delta t\coloneqq t\ts{B,0}-t\ts{A,0}$. 
In $(3+1)$-dimensional Minkowski spacetime, the detectors can potentially communicate when they are lightlike separated. 
This is depicted as an orange strip in the figure; detector-A  can send quanta to detector-B once the support of detector-B has overlap with this region.

We first ask if  initially entangled detectors can gain more entanglement after interacting with their local fields? 
The answer is \textit{yes}, though we find that the condition is very limited; only initially weakly entangled detectors can gain entanglement. 

Consider weakly entangled detectors with $\beta \approx 0$ (i.e., $\alpha \approx 1$). 
Using~\eqref{eq:approx r14 alpha 1} to calculate the concurrence $\mathcal{C}\ts{AB}$ after the interaction, we illustrate in
Figure~\ref{fig:spacetime diagram and plot}(b)   the difference between the initial and final concurrences of the detectors for $\beta=10^{-15}, \lambda=1/10, \Omega T=5, \sigma/T=0$, and $\theta=0$. 
In particular, we take $\max \{ 0,\,\mathcal{C}\ts{AB}-\mathcal{C}\ts{AB,0} \}$ so that we can see the region of entanglement enhancement. We depict with green dots values
of  the concurrence that are less than its initial value $\mathcal{C}\ts{AB,0}$  after the interaction. 
We first note that communication between the detectors greatly assists entanglement extraction as one can see from Fig.~\ref{fig:spacetime diagram and plot}(b) that two detectors gain entanglement when they are in causal contact (within the dashed lines). 
This phenomenon can be explained from the entanglement harvesting viewpoint; communication enhances the value of $|X\ts{AB}(-\Omega)|$ in the entanglement harvesting scenario ($\beta=0$) \cite{TjoaSignal}, and this is also happening in our $|r_{14}|$ in \eqref{eq:approx r14 alpha 1}. 
Nevertheless, as described in the previous section, the degradation term $\text{Re}[Y\ts{A}+Y\ts{B}]$ will suppress this communication assistance if the interaction duration is too long.

One can also ask if noncommunicating detectors can gain entanglement from the field. 
The answer is also \textit{yes}. 
To see this, we plot concurrence for pointlike detectors ($\sigma/T=0$) when they are effectively\footnote{Since we are using a Gaussian switching, the interaction is not compactly supported in the spacetime. 
In other words, the detectors are always communicating. 
However, due to a Gaussian function's exponential suppression, it can be thought of as an effectively compactly supported interaction in $\tau \in [ -3.5 T + \tau_{j,0}, 3.5 T + \tau_{j,0} ]$, where $\tau$ is a proper time and $\tau_{j,0}$ is the center of Gaussian of detector-$j$ \cite{TjoaSignal}.} causally disconnected ($L/T=7, \Delta t/T=0$) in Fig.~\ref{fig:Omega and beta plot}. 
Figure~\ref{fig:Omega and beta plot}(a) shows the values of $\Omega T$ and $\log_{10}\beta$ that enhance entanglement without communication. 
The parameters are set to be $\lambda=1/10, L/T=7, \Delta t/T=0,  \sigma/T=0$, and $\theta=0$. 
Again, the detectors experience degradation when $(\Omega T, \log_{10}\beta)$ is chosen to be a point in the green dots in Figure~~\ref{fig:Omega and beta plot}(a). 
 We find that a pair of causally disconnected detectors can enhance their initial entanglement as long as they are weakly entangled at the beginning. 
In addition, one must choose $\Omega T$ from a suitable range: if the initial entanglement is, for example,  $\log_{10}\beta=-30$ with the given parameters, then the energy gap must be $\Omega T \in [6.8, 11]$ to gain more correlation. 
This energy gap range increases as $\beta \to 0$, namely, the upper bound of $\Omega$ to extract entanglement gets pushed to infinity in the limit $\beta \to 0$. 
We also note that there is a maximum value of $\beta$ for fixed $L/T$ beyond which entanglement cannot be extracted for any $\Omega T$. 

The blue curve in Fig.~\ref{fig:Omega and beta plot}(b) represents a slice at $\Omega T=7$ in (a), and each factor $\mathcal{C}\ts{AB}$ and $\mathcal{C}\ts{AB,0}$ in $\max\{ 0,\,\mathcal{C}\ts{AB}-\mathcal{C}\ts{AB,0} \}$ is shown in (c) as blue and green curves, respectively. 
It is interesting to note from (c) that for sufficiently large $\beta$, entanglement is completely extinguished after the interaction. 
There is also a dependence on the relative initial phase $\theta$; for $\theta=\pi$ [the orange curve in (b) and (c)], the positive difference $\mathcal{C}\ts{AB}-\mathcal{C}\ts{AB,0}$ persists for larger values of $\beta$, allowing for a wider range of entanglement extraction. 
This dependence on the relative phase $\theta$ arises from the term $\text{Re}[ X\ts{AB}^*(-\Omega) e^{-\ii \theta} ]$ in \eqref{eq:approx r14 alpha 1} and elements in $r_{22} r_{33}$.

For the case of initially sufficiently entangled detectors, no entanglement extraction can occur --- only degradation takes place. 
One of the reasons comes from the fact that the degradation part $\text{Re}[Y\ts{A} + Y\ts{B}]$ is no longer suppressed in \eqref{eq:approx r14 alpha 1} as $\beta$ grows. 
Extracting entanglement with $\alpha \approx 0$ is also difficult since $I_{kk}^{(+-)}$ in $\sqrt{r_{22} r_{33}}$ dominates with large $\Omega$.

\begin{figure}[t]
    \centering
    \includegraphics[width=\columnwidth]{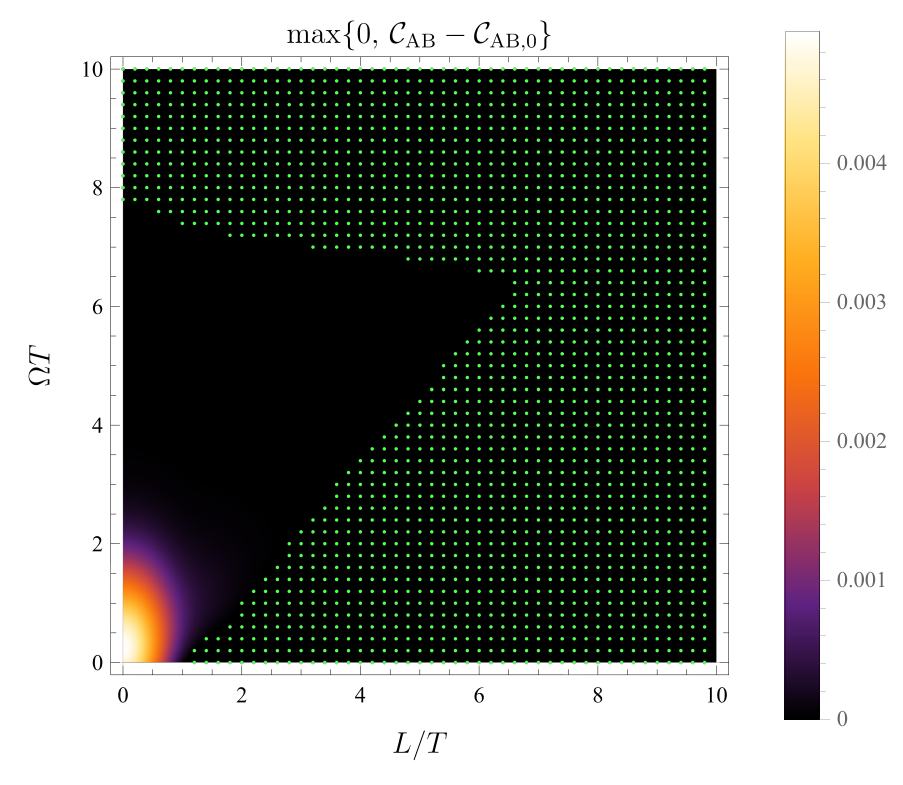}\\
    \caption{
    ~The positive part of difference $\mathcal{C}\ts{AB}-\mathcal{C}\ts{AB,0}$ as a function of $L/T$ and $\Omega T$. 
    Here, $\Delta t/T=0, \beta= 10^{-15}$, and $\theta=0$. 
    Entanglement degradation occurs in the region with green dots.}
    \label{fig:ConcDiffLandOmega}
\end{figure}

One notable difference from entanglement harvesting ($\beta=0$) is the energy gap dependence. 
In the harvesting scenario, detectors with a Gaussian switching\footnote{This is not the case for sudden switchings \cite{pozas2015harvesting}.} can extract entanglement even when they are far apart ($L/T \gg 1$) if $\Omega $ is large enough \cite{pozas2015harvesting}. 
That is, once the detector separation $L$ is fixed, there exists a minimum energy gap $\Omega\ts{min}$ such that the detectors can harvest entanglement for all $\Omega \geq \Omega\ts{min}$. 
In our case, however, this is no longer true. 
Figure \ref{fig:ConcDiffLandOmega} shows $\max \{0,\, \mathcal{C}\ts{AB} - \mathcal{C}\ts{AB,0}\}$ as a function of $L/T$ and $\Omega T$ for pointlike detectors with $\Delta t=0, \beta=10^{-15}$, and $\theta=0$. 
That is, we are exploring the region of entanglement extraction in $(L/T, \Omega T)$-plane when the centers of Gaussian switching in Fig.~\ref{fig:spacetime diagram and plot}(a) are on the same time-slice. 
We see that there exist minimum and maximum values of $\Omega$ for some $L$ that allow detectors to extract entanglement;  once $\Omega$ gets large enough, only degradation takes place. 
The maximum energy gap  $\Omega\ts{max}$ allowing harvesting is pushed to infinity as $\beta \to 0$,  recovering the harvesting scenario   \cite{pozas2015harvesting}.

\section{Entanglement in the presence of a shockwave}
\label{sec:UDW detectors in presence of a shockwave}

We now extend our results in Minkowski spacetime to a nontrivial spacetime. We consider in particular a gravitational shockwave spacetime, since it has been shown   that a weak gravitational field can enhance entanglement harvesting from the vacuum \cite{Cliche.weak.gravity}. Furthermore, for the type of gravitational shockwave that we consider in this paper, the Wightman function for a massless scalar field   has a non-trivial, but closed form expression~\cite{FinnShockwave}. 

This latter study considered initially separable detectors in their ground states; we consider here initially entangled detectors with the aim of understanding if the shockwave mitigates or enhances entanglement degradation.
We find that a shockwave admits for greater entanglement harvesting relative to the flat space setting, and extends the region in Fig.~\ref{fig:Omega and beta plot}(a) dramatically.

\subsection{Shockwave spacetime}
\label{subsec:Shockwave spacetime}

As in \cite{FinnShockwave} we shall be working with the Dray and 't Hooft generalization \cite{Dray.shockwave} of the Aichelburg-Sexl shockwave spacetime \cite{Aichelburg1971gravitational}. 
Consider a planar shockwave propagating in the $z$-direction in $D$-dimensional Minkowski spacetime. 
The metric describing such a spacetime in the so-called Brinkmann coordinates \cite{Brinkmann1923riemann} is:
\begin{equation}
    \dd s^2
    =
        -\dd u \dd v + f(\vec{x}) \delta(u-u_0) \dd u^2
        + \delta_{ij}\dd x^i \dd x^j\,,
\end{equation}
where $u=t-z$ and $v=t+z$ are null coordinates, and $\vec{x}$ or $x^i$, $i\in \{ 1, \cdots, D-2 \}$ are the transverse coordinates (i.e., the remaining spatial coordinates). 
The wavefront of the shockwave is located at $u=u_0$. 
We consider an idealized scenario in which the spacetime is exactly Minkowski on either side of $u_0$.

\begin{figure}[b]
    \centering
    \includegraphics[width=7cm]{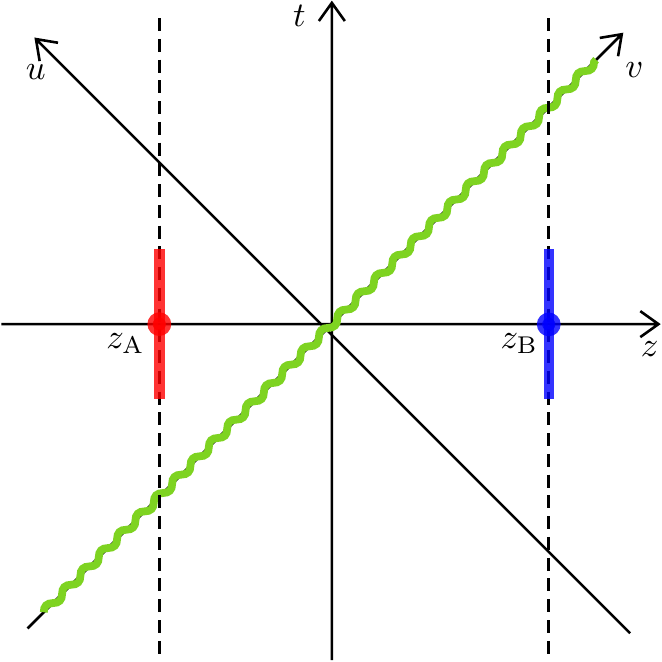}\\
    \caption{
    ~Diagram for two pointlike inertial detectors in a shockwave spacetime. 
    Detectors A and B are static at $z=z\ts{A}$ and $z\ts{B}$, respectively, and the shockwave travels along $u=0$ (the green wiggling line). 
    The red and blue bars indicate the effective interaction duration of the detectors. }
    \label{fig:shockwave diagram}
\end{figure}

The term $f(\vec{x})$ is the shockwave profile, which is directly related to the energy density $\varrho(\vec{x})$ through the Einstein field equations: 
\begin{align}
    \Delta f(\Vec{x})
    &=
        -16 \pi G\ts{N} \varrho(\Vec{x})\,, \label{eq:EFE for shockwave}
\end{align}
where $\Delta=\delta^{ij}\partial_i\partial_j$ is the flat Laplacian in the transverse direction. 
We use a quadratic profile given by
\begin{equation}
    f(\vec{x})=-\vec{x}\cdot A\cdot\vec{x}=-\sum_{i=1}^{D-2}a_{i}(x^i)^2, 
\end{equation}
where $A$ is a symmetric constant matrix, whose eigenvalues are $a_i$. 
Such a profile represents an ultrarelativistic domain wall in the transverse plane \cite{Lousto.domainwall} as one can see from \eqref{eq:EFE for shockwave} that the energy density is constant everywhere: $\varrho(\Vec{x})=\Tr[A]/8\pi G\ts{N}$.

\begin{figure*}[t]
    \centering
    \includegraphics[width=\textwidth]{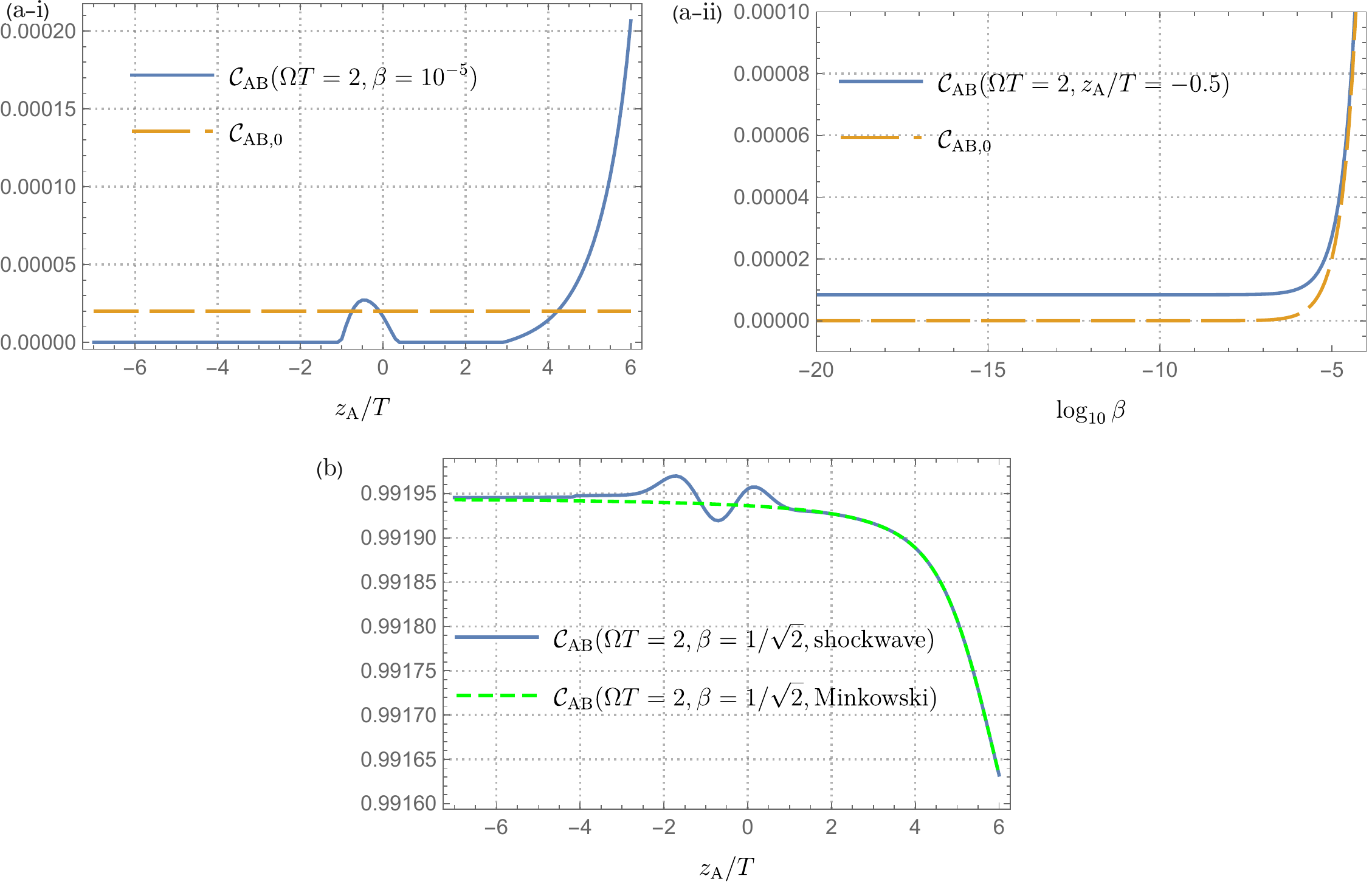}\\
    \caption{
    (a-i) Initial and final concurrences when $\Omega T=2, \beta=10^{-5}, \lambda=1/10$, and $\theta=0$. 
    The location of detector-B is fixed at $z\ts{B}/T=7$. 
    Detector-A encounters the shockwave near $z\ts{A}/T=0$. 
    (a-ii) The concurrences when $\Omega T=2, z\ts{A}/T=-0.5$ and $z\ts{B}/T=7$ with varying initial amount of entanglement. 
    Two curves will cross and degradation takes place after $\log_{10}\beta >- 3$. 
    (b) Final concurrence $\mathcal{C}\ts{AB}$ for initially maximally entangled detectors ($\mathcal{C}\ts{AB,0}=1$) with $\Omega T=2$ and $\theta=0$. 
    The green dotted curve corresponds to detectors in Minkowski spacetime. 
    In this case, the shockwave can either reduce or amplify the degradation effect. 
   }
    \label{fig:Shockwave Fig}
\end{figure*}

Using the notation $\sx=(u,v,\vec{x})$ and $\sy=(U,V,\vec{X})$   the exact form of the Wightman function is \cite{FinnShockwave}
\begin{widetext}
\begin{equation}
    \begin{split}
        W(\sx,\sy)
        &=
            \frac{(-\ii)^{D-2}\Gamma(D/2-1)}{4\pi^{D/2}}
            \prod_{i=1}^{D-2}
            \bigg(
                1 + a_i \Delta\Theta
                \frac{(u-u_0 - \ii\epsilon)(U-u_0 + \ii\epsilon)}{\Delta u- \ii\epsilon}
            \bigg)^{-1/2} \\
            &\quad \times
            \bigg(
                (\Delta v- \ii\epsilon)
                (\Delta u-\ii\epsilon) 
                -\Delta\vec{x}^2 
                + \sum_{i=1}^{D-2} \frac{a_i\Delta \Theta \big([U-u_0+ \ii \epsilon]x^i -[u-u_0- \ii\epsilon] X^i\big)^2}{\Delta u- \ii\epsilon + a_i \Delta\Theta (u-u_0-\ii\epsilon)(U-u_0 + \ii\epsilon)}
            \bigg)^{(2-D)/2}\,,
    \end{split}
\end{equation}
\end{widetext}
where $\Delta v\equiv v-V$, $\Delta u \equiv u-U$, $\Theta_u\equiv\Theta(u-u_0)$, and $\Delta\Theta=\Theta_u-\Theta_U$ with the Heaviside step function $\Theta(u)$. 
$\Gamma(x)$ is the gamma function and by choosing $D=4$ it gives $\Gamma(1)=1$. 
The Wightman function reduces to the standard Minkowski space form if $f(\vec{x})=0$ (in which case there is no shockwave at all) or $\Delta\Theta=0$ (so the events $u$ and $U$ are localized at the same side of the shockwave).

\subsection{Results}
\label{sec: results}

We restrict to $D=4$ and consider two inertial UDW detectors 
\begin{equation}
    \begin{split}
        &\sx\ts{A}(\tau\ts{A})
        =
            (t(\tau\ts{A}),x(\tau\ts{A}),y(\tau\ts{A}),z(\tau\ts{A}))
        =
            (\tau\ts{A}, 0, 0, z\ts{A})\,, \\
        &\sx\ts{B}(\tau\ts{B})
        =
            (t(\tau\ts{B}), x(\tau\ts{B}), y(\tau\ts{B}), z(\tau\ts{B}))
        =
            (\tau\ts{B}, 0, 0, z\ts{B})\,.
    \end{split}
\end{equation}
We further assume that $a_x=a_y \equiv a (>0)$ and $u_0=0$ as in Fig.~\ref{fig:shockwave diagram} and $\vec{x}=\vec{X}=\vec{0}$. 
The Wightman function then reduces to 
\begin{align}
    &W(\sx, \sy)
    = \notag \\
        &-\dfrac{1}{4\pi^2}
        \dfrac{1}{ 1 + a \Delta \Theta \dfrac{ (u-\ii \epsilon) (U+\ii \epsilon) }{ \Delta u -\ii \epsilon } }
        \dfrac{1}{ (\Delta v -\ii \epsilon) (\Delta u -\ii \epsilon) }\,.
\end{align}
It is worth noting that in this setup, the proper spatial distance between the two detectors, $L=|z\ts{A} - z\ts{B}|$, is the same as in Minkowski space, and that the two detectors are placed in the longitudinal direction of the shockwave.

% Also as in \cite{FinnShockwave} the switching functions that specify how strongly the detector interacts with the scalar field over its proper time are Gaussians with width $T$:

% \begin{equation}
%     \chi(\tau)=\exp{\left[-\frac{(\tau-\tau_0)^2}{T^2}\right]}\,.
% \end{equation}

In Fig.~\ref{fig:Shockwave Fig}(a), we plot the initial ($\mathcal{C}\ts{AB,0}$)  and final ($\mathcal{C}\ts{AB}$) concurrences in the shockwave spacetime with $aT=1$. 
In all figures, detector-B is fixed at the position $z\ts{B}/T=7$ which is located on the $u<u_0$ part of shockwave spacetime (see Fig.~\ref{fig:shockwave diagram}) and the energy gap is $\Omega T=2$.
Note that in Minkowski space, the noncommunicating detectors cannot extract entanglement with $\Omega T=2$, as shown in Figs.~\ref{fig:Omega and beta plot}(a) and \ref{fig:ConcDiffLandOmega}. 

Figure~\ref{fig:Shockwave Fig}(a-i) depicts the concurrences as functions of detector-A's static positions $z\ts{A}/T$ when $\beta=10^{-5}$. 
As shown in Fig.~\ref{fig:shockwave diagram}, detector-A encounters the shockwave around $z\ts{A}/T=0$. 
For $z\ts{A}/T \lesssim -1$ the plot indicates that full entanglement degradation from the initial value of $\mathcal{C}\ts{AB,0}$ occurs, due to the fact that the geometry is Minkowski except along the shockwave trajectory. 
In other words, the behavior in $z\ts{A}/T \lesssim -1$ follows from the results in Minkowski spacetime given in Figs.~\ref{fig:Omega and beta plot} and \ref{fig:ConcDiffLandOmega}. 
This is also true for $z\ts{A}/T \gtrsim 0.5$. 
The growth of $\mathcal{C}\ts{AB}$ in $z\ts{A}/T > 3$ comes from the communication effect in $|X\ts{AB}(-\Omega)|$ between the two detectors, which corresponds to the bottom left corner in Fig.~\ref{fig:spacetime diagram and plot}(b). 
The effect of the shockwave can be seen in the region $z\ts{A}/T \in [-1, 0.5]$; not only is the degradation effect reduced, but we find $\mathcal{C}\ts{AB}>\mathcal{C}\ts{AB,0}$ ---
the initial entanglement is enhanced!
Somewhat surprisingly, the shockwave allows causally disconnected detectors to gain more entanglement even when it is impossible to do this in Minkowski spacetime.

Figure~\ref{fig:Shockwave Fig}(a-ii) shows the $\beta$-dependence of the concurrences when $z\ts{A}/T=-0.5$ (i.e., $L/T=7.5$), where maximal entanglement enhancement occurs, with $\Omega T=2, \lambda=1/10$, and $\theta=0$. 
We find that entanglement enhancement by the shockwave is possible so long as $\beta \lesssim 10^{-3}$. 
Recall that for any $\beta$, only degradation occurs for these parameters in Minkowski spacetime. 
Furthermore, even when $\Omega T=7$ as in Figs.~\ref{fig:Omega and beta plot}(b) and (c), the initial amount of entanglement needs to be $\beta \lesssim 10^{-17}$ in order to extract entanglement. 
In this sense, a shockwave drastically assists extraction of entanglement.

However, the shockwave does not always enhance entanglement. 
In Fig.~\ref{fig:Shockwave Fig}(b), we plot the final concurrence when the detectors are initially maximally entangled, $\mathcal{C}\ts{AB,0}=1$. 
In both Minkowski (dotted curve) and shockwave (solid) spacetimes, degradation takes place as expected. 
Notice that the shockwave could either reduce or assist degradation. 
This is due to the fact that the concurrence is now determined by \eqref{eq:approx r14 sufficient}; the degradation part $\text{Re}[Y\ts{A}+Y\ts{B}]$ is no longer suppressed by small $\beta^2$ and the harvesting part $|X\ts{AB}(-\Omega)|$ is negligible.

\section{Conclusion}
\label{sec: conclusion}

We considered two initially entangled UDW detectors interacting with a quantum scalar field and examined how their initial entanglement behaves after the interaction. The entanglement harvesting protocol allows initially uncorrelated detectors to extract entanglement from the field in its vacuum state without signalling, whereas entanglement degradation is the process in which initial entanglement decreases due to the detectors becoming entangled with the field. 
We have investigated which phenomenon is dominant  when the detectors are neither uncorrelated nor maximally entangled.

Such an analysis is known to be difficult when a perturbative method is employed since there is a UV-divergent element in the density matrix for the pointlike detectors (and hence in entanglement measures). 
One can circumvent this issue by introducing a smearing function and performing an approximation to an entanglement measure such as concurrence and negativity. 
We have dealt with this by introducing an approximation for weakly entangled detectors, supplementing the method used for initially sufficiently entangled detectors~\cite{Chowdhury.FateEntanglement}. 
Under any of these approximations, the divergent element is absent in the entanglement measure.

With these approximations, we examined the fate of entanglement in detectors at rest in $(3+1)$-dimensional Minkowski spacetime. 
We found that \textit{initially weakly entangled detectors can gain entanglement even when they are causally disconnected}. 
This can be understood by looking at the distinct contributions to the concurrence: one part is from the initial entanglement, another yields entanglement harvesting, and a third causes entanglement degradation. 
The detectors extract entanglement if the entanglement harvesting contribution to  the concurrence exceeds the other two. 
Otherwise, entanglement degradation takes place and the initial amount of entanglement will be reduced after the interaction. 

The fact that only weakly entangled detectors can gain entanglement is consistent with the results in Ref.~\cite{GallockEntangledDetectors}. 
It was nonperturbatively shown that quantum mutual information (i.e., the total correlation including classical ones) between entangled detectors can be enhanced only when their initial entanglement is weak\footnote{The nonperturbative method employed in \cite{GallockEntangledDetectors} uses a delta-switching function, which cannot harvest entanglement at all if each detector switches only once \cite{Simidzija.Nonperturbative, Simidzija2018no-go}}. 
It is important to note, however, that the detectors can gain mutual information even when they are causally disconnected.

We have also analyzed the energy gap $\Omega$ dependence for entanglement extraction. 
In the entanglement harvesting scenario \cite{pozas2015harvesting}, detectors with Gaussian switching can extract entanglement so long as their energy gap is above the minimum value: $\Omega \geq \Omega\ts{min}$. 
This feature allows one to harvest entanglement with detectors arbitrarily far away from each other once a sufficiently large energy gap is chosen. 
This is not the case for entangled detectors. 
Instead there is a range $\Omega \in [\Omega\ts{min}, \Omega\ts{max}]$ within which the detectors can  extract entanglement. 
This range tends to zero as the distance between the detectors increases. 
This suggests that entanglement extraction is not allowed for detectors arbitrarily far away from each other. 
Nevertheless, the maximum value $\Omega\ts{max}$ will increase as  the initial state approaches a separable state,  recovering the results of the standard entanglement harvesting protocol. 

We found that the presence of a gravitational shockwave, in which the metric is identical to that of Minkowski spacetime except along a single null trajectory, significantly modifies these effects.  
Detectors far from the shockwave exhibit   the same behavior as in Minkowski spacetime. 
However as they come closer to the shockwave their behavior markedly changes.
The shockwave enhances entanglement as compared to Minkowski spacetime. 
The range  $\Omega \in [\Omega\ts{min}, \Omega\ts{max}]$ of energy gap that enables entanglement extraction becomes wider, which in turn allows the detectors to gain entanglement for smaller values of $\Omega$. 

Although the shockwave enhances entanglement, its extraction is still limited to weakly entangled detectors. 
If the initial state of the detectors is sufficiently entangled, the shockwave weakens the effect of degradation rather than enhancing extraction of correlations. 
Nevertheless, our results show that  spacetime geometry can enhance initial entanglement harvesting, indicating that the quantum vacuum is indeed a resource (at least in principle) for carrying out quantum information tasks.

We close by commenting on a recent similar investigation of the negativity~\cite{Barman.spontaneous} of  two initially entangled detectors in Minkowski spacetime with constant switching. 
Using an approximation corresponding to our sufficiently entangled case, the detectors were found to  only experience entanglement degradation, consistent with our results for sufficiently entangled detectors with an infinitely long interaction.

\section*{Acknowledgments}

This work was supported in part by the Natural Sciences and Engineering Research Council of Canada.

\begin{widetext}

\appendix
\section{Smeared detectors in Minkowski spacetime}
\label{app:smeared detectors}
The imaginary part of $\lambda^2 Y_k \equiv J_{kk}^{(-+)}+J_{kk}^{(+-)*}$ in $r_{14}$ for pointlike detectors is known to be divergent. 
We will show that this term can be regularized by introducing finite-sized UDW detectors as suggested in \cite{EMM.Jorma.Firewall}. 
In particular, we consider initially entangled smeared detectors at rest in $(3+1)$-dimensional Minkowski spacetime.

Assuming a massless scalar field, the mode decomposition of $\hat \phi(\sx(t))$ is known to be 
\begin{align}
    \hat \phi(t, \bm{x})
    &=
        \int \dfrac{ \dd^3 k }{ \sqrt{ (2\pi)^3 2 \kk } }
        \kako{
            \hat a_{\bm{k}} e^{ -\ii \kk t + \ii \bm{k}\cdot \bm{x} }
            +
            \hat a_{\bm{k}}^\dag e^{ \ii \kk t - \ii \bm{k}\cdot \bm{x} }
        },
\end{align}
and so the Wightman function becomes 
\begin{align}
    W(\sx_j(t_1) , \sy_k(t_2))
    &=
        \int \dfrac{ \dd^3k }{ (2\pi)^3 2\kk }
        e^{ -\ii \kk (t_1-t_2) + \ii \bm{k}\cdot (\bm{x} - \bm{y}) }. 
\end{align}

\begin{figure}[t]
    \centering
    \includegraphics[width=10cm]{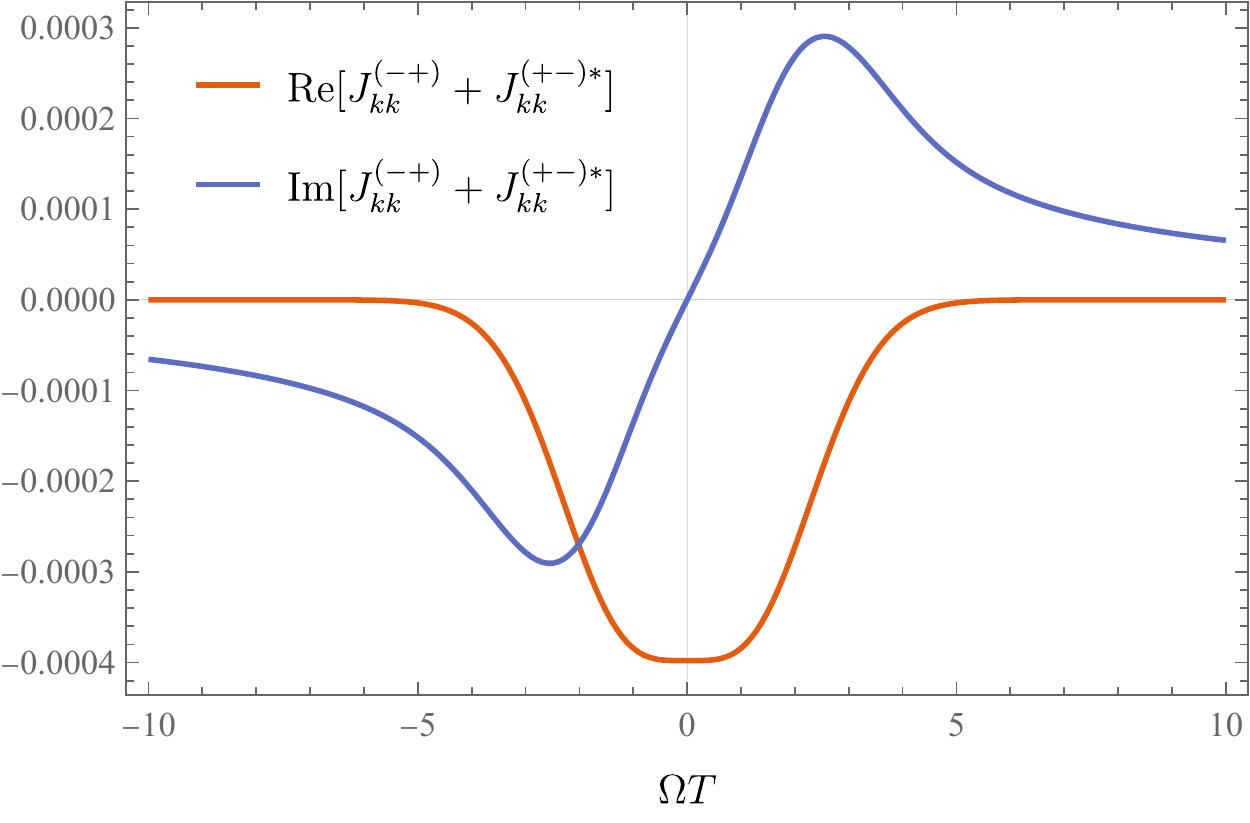}\\
    \caption{
    ~Real and imaginary parts of $J_{kk}^{(-+)}+J_{kk}^{(+-)*}$ with $\sigma /T=1$ and $\lambda = 1/10$. 
    The real part converges quickly to 0 with large $|\Omega T|$ while the imaginary part is nonzero for a wide range of $\Omega T$. }
    \label{fig:divergent element}
\end{figure}

Let us first consider $\lambda^2 Y\ts{A} \equiv J\ts{AA}^{(-+)}+J\ts{AA}^{(+-)*}$ for detector-A in $r_{14}$.  We find
\begin{align}
    &J\ts{AA}^{(-+)}+J\ts{AA}^{(+-)*} \notag \\
    &=
        -2\lambda\ts{A}^2 
        \int_{\mathbb{R}} \dd t_1
        \int_{-\infty}^{t_1} \dd t_2\,
        \chi\ts{A}(t_1)
        \chi\ts{A}(t_2)
        e^{ -\ii \Omega\ts{A} (t_1 - t_2) } 
        \int \dd^3 x\,F\ts{A}(\bm{x}-\bm{x}\ts{A})
        \int \dd^3 y\,F\ts{A}(\bm{y}-\bm{x}\ts{A})
        \text{Re}[ W(\sx\ts{A}(t_1), \sy\ts{A}(t_2)) ] \\
    &=
        -2\lambda\ts{A}^2 
        \int \dfrac{ \dd^3k }{ 2\kk }
        \tilde F\ts{A}(\bm{k}) \tilde{F}\ts{A}(-\bm{k})
        \int_{\mathbb{R}} \dd t_1
        \int_{-\infty}^{t_1} \dd t_2\,
        \chi\ts{A}(t_1)
        \chi\ts{A}(t_2)
        e^{ -\ii \Omega\ts{A} (t_1 - t_2) } 
        \cos [ \kk (t_1-t_2) ] \,,
\end{align}
where we have performed the Fourier transform: 
\begin{align}
    \tilde F_j (\bm{k})
    &=
        \int \dfrac{ \dd^3 x }{ \sqrt{ (2\pi)^3 } }
        F_j(\bm{x}) e^{ \ii \bm{k}\cdot \bm{x} }\,. 
\end{align}
Note that $\tilde F_j(-\bm{k})=\tilde F_j(\bm{k})$ if $F_j(\bm{x}) \in \mathbb{R}$. 
With a Gaussian switching centered at $t=t\ts{A,0}$, $\chi\ts{A}(t)=e^{ -(t-t\ts{A,0})^2/T^2 }$, 
\begin{align}
    &\int_{\mathbb{R}} \dd t_1
        \int_{-\infty}^{t_1} \dd t_2\,
        e^{ -(t_1-t\ts{A,0})^2/T^2 }
        e^{ -(t_2-t\ts{A,0})^2/T^2 }
        e^{ -\ii \Omega\ts{A} (t_1 - t_2) } 
        \cos [ \kk (t_1-t_2) ] \notag \\
    &=
        \dfrac{T\sqrt{2\pi}}{2} 
        \int_0^\infty \dd u\,
        e^{ -u^2/2T^2 } e^{ -\ii \Omega u }
        \cos ( \kk u ) \\
    &=
        \dfrac{\pi T^2}{4}
        e^{ -T^2(\kk+\Omega)^2/2 }
        \kagikako{
            1 + 
            e^{ 2\kk T^2 \Omega } 
            \kako{
                1 + \ii \erfi \dfrac{ T(\kk - \Omega) }{\sqrt{2} }
            }
            -\ii \erfi \dfrac{ T(\kk + \Omega) }{\sqrt{2} }
        }, 
\end{align}
and therefore, using the smearing function \eqref{eq:smearing function} and its Fourier transform, $\tilde{F}_j(\bk)=e^{ -\kk^2 \sigma^2/4 }/\sqrt{ (2\pi)^{3} }$,
\begin{align}
    &J\ts{AA}^{(-+)} + J\ts{AA}^{(+-)*} \notag \\
    &=
        -\dfrac{\lambda\ts{A}^2 \pi T^2}{2}
        \int \dfrac{ \dd^3k }{ 2\kk }
        \tilde F\ts{A}^2(\bm{k}) 
        e^{ -T^2(\kk+\Omega)^2/2 } 
        \kagikako{
            1 + 
            e^{ 2\kk T^2 \Omega } 
            \kako{
                1 + \ii \erfi \dfrac{ T(\kk - \Omega) }{\sqrt{2} }
            }
            -\ii \erfi \dfrac{ T(\kk + \Omega) }{\sqrt{2} }
        } \\
    &=
        -\dfrac{ \lambda\ts{A}^2 e^{ -T^2 \Omega^2/2 } }{ 8\pi (1 + \sigma^2/T^2)^{3/2} }
        \kagikako{
            2 \sqrt{1 + \sigma^2/T^2}
            +
            e^{ T^2 \Omega^2/2(1 + \sigma^2/T^2) }
            \sqrt{2\pi} T \Omega 
            \erf \dfrac{ T \Omega }{ \sqrt{ 2(1 + \sigma^2/T^2) } }
        } \notag \\
        &\quad
        - \ii 
        \dfrac{ \lambda\ts{A}^2 T^2 }{8\pi }
        \int_0^\infty \dd \kk\,
        \kk 
        e^{ -\kk^2 \sigma^2/2 }
        \kagikako{
            e^{ -T^2(\kk-\Omega)^2/2 } 
            \erfi \dfrac{ T(\kk - \Omega) }{\sqrt{2} }
            - 
            e^{ -T^2(\kk+\Omega)^2/2 } 
            \erfi \dfrac{ T(\kk + \Omega) }{\sqrt{2} }
        }\,,
\end{align}
where we have used $\dd^3k = \kk^2 \sin \theta \dd \kk \dd \theta \dd \varphi$ with $\kk \in [0,\infty), \theta\in [ 0, \pi ]$, and $\varphi\in [0,2\pi)$. 
The real and imaginary parts of $J_{kk}^{(-+)}+J_{kk}^{(+-)*}$ for smeared detectors are shown in Fig.~\ref{fig:divergent element}.

Similar calculations yield   the remaining terms in $r_{14}, r_{22}$, and $r_{33}$. 
In particular, the expressions for $I_{kk}^{(-+)}$ in $r_{22}$ 
and
 $r_{33}$   in \eqref{eq:Ikkpm}
and  for $J\ts{AB}^{(++)}$ (and its counterparts)
in \eqref{eq:JABmm} that contribute to
 $\lambda^2 X\ts{AB}^*(-\Omega)\equiv J\ts{AB}^{(++)*}+J\ts{BA}^{(++)*}$ in $r_{14}$ can be found in Ref.~\cite{pozas2015harvesting}.

\end{widetext}

\bibliography{ref}

\begin{thebibliography}{45}%
\makeatletter
\providecommand \@ifxundefined [1]{%
 \@ifx{#1\undefined}
}%
\providecommand \@ifnum [1]{%
 \ifnum #1\expandafter \@firstoftwo
 \else \expandafter \@secondoftwo
 \fi
}%
\providecommand \@ifx [1]{%
 \ifx #1\expandafter \@firstoftwo
 \else \expandafter \@secondoftwo
 \fi
}%
\providecommand \natexlab [1]{#1}%
\providecommand \enquote  [1]{``#1''}%
\providecommand \bibnamefont  [1]{#1}%
\providecommand \bibfnamefont [1]{#1}%
\providecommand \citenamefont [1]{#1}%
\providecommand \href@noop [0]{\@secondoftwo}%
\providecommand \href [0]{\begingroup \@sanitize@url \@href}%
\providecommand \@href[1]{\@@startlink{#1}\@@href}%
\providecommand \@@href[1]{\endgroup#1\@@endlink}%
\providecommand \@sanitize@url [0]{\catcode `\\12\catcode `\$12\catcode
  `\&12\catcode `\#12\catcode `\^12\catcode `\_12\catcode `\%12\relax}%
\providecommand \@@startlink[1]{}%
\providecommand \@@endlink[0]{}%
\providecommand \url  [0]{\begingroup\@sanitize@url \@url }%
\providecommand \@url [1]{\endgroup\@href {#1}{\urlprefix }}%
\providecommand \urlprefix  [0]{URL }%
\providecommand \Eprint [0]{\href }%
\providecommand \doibase [0]{https://doi.org/}%
\providecommand \selectlanguage [0]{\@gobble}%
\providecommand \bibinfo  [0]{\@secondoftwo}%
\providecommand \bibfield  [0]{\@secondoftwo}%
\providecommand \translation [1]{[#1]}%
\providecommand \BibitemOpen [0]{}%
\providecommand \bibitemStop [0]{}%
\providecommand \bibitemNoStop [0]{.\EOS\space}%
\providecommand \EOS [0]{\spacefactor3000\relax}%
\providecommand \BibitemShut  [1]{\csname bibitem#1\endcsname}%
\let\auto@bib@innerbib\@empty
%</preamble>
\bibitem [{\citenamefont {Summers}\ and\ \citenamefont
  {Werner}(1985)}]{summers1985bell}%
  \BibitemOpen
  \bibfield  {author} {\bibinfo {author} {\bibfnamefont {S.~J.}\ \bibnamefont
  {Summers}}\ and\ \bibinfo {author} {\bibfnamefont {R.}~\bibnamefont
  {Werner}},\ }\bibfield  {title} {\bibinfo {title} {{The vacuum violates
  Bell's inequalities}},\ }\href
  {https://doi.org/https://doi.org/10.1016/0375-9601(85)90093-3} {\bibfield
  {journal} {\bibinfo  {journal} {Phys. Lett.}\ }\textbf {\bibinfo {volume}
  {110A}},\ \bibinfo {pages} {257 } (\bibinfo {year} {1985})}\BibitemShut
  {NoStop}%
\bibitem [{\citenamefont {Summers}\ and\ \citenamefont
  {Werner}(1987)}]{summers1987bell}%
  \BibitemOpen
  \bibfield  {author} {\bibinfo {author} {\bibfnamefont {S.~J.}\ \bibnamefont
  {Summers}}\ and\ \bibinfo {author} {\bibfnamefont {R.}~\bibnamefont
  {Werner}},\ }\bibfield  {title} {\bibinfo {title} {Bell’s inequalities and
  quantum field theory. i. general setting},\ }\href
  {https://doi.org/10.1063/1.527733} {\bibfield  {journal} {\bibinfo  {journal}
  {J. Math. Phys. (N.Y.)}\ }\textbf {\bibinfo {volume} {28}},\ \bibinfo {pages}
  {2440} (\bibinfo {year} {1987})}\BibitemShut {NoStop}%
\bibitem [{\citenamefont {Unruh}(1976)}]{Unruh1979evaporation}%
  \BibitemOpen
  \bibfield  {author} {\bibinfo {author} {\bibfnamefont {W.~G.}\ \bibnamefont
  {Unruh}},\ }\bibfield  {title} {\bibinfo {title} {Notes on black-hole
  evaporation},\ }\href {https://doi.org/10.1103/PhysRevD.14.870} {\bibfield
  {journal} {\bibinfo  {journal} {Phys. Rev. D}\ }\textbf {\bibinfo {volume}
  {14}},\ \bibinfo {pages} {870} (\bibinfo {year} {1976})}\BibitemShut
  {NoStop}%
\bibitem [{\citenamefont {{DeWitt}}(1979)}]{DeWitt1979}%
  \BibitemOpen
  \bibfield  {author} {\bibinfo {author} {\bibfnamefont {B.~S.}\ \bibnamefont
  {{DeWitt}}},\ }\bibfield  {title} {\bibinfo {title} {{Quantum gravity: The
  new synthesis}},\ }in\ \href@noop {} {\emph {\bibinfo {booktitle} {General
  Relativity: An Einstein Centenary Survey}}},\ \bibinfo {editor} {edited by\
  \bibinfo {editor} {\bibfnamefont {S.~W.}\ \bibnamefont {{Hawking}}}\ and\
  \bibinfo {editor} {\bibfnamefont {W.}~\bibnamefont {{Israel}}}}\ (\bibinfo
  {year} {1979})\ pp.\ \bibinfo {pages} {680--745}\BibitemShut {NoStop}%
\bibitem [{\citenamefont {Mart\'{\i}n-Mart\'{\i}nez}\ \emph
  {et~al.}(2013)\citenamefont {Mart\'{\i}n-Mart\'{\i}nez}, \citenamefont
  {Montero},\ and\ \citenamefont {del Rey}}]{EMM.wavepacket}%
  \BibitemOpen
  \bibfield  {author} {\bibinfo {author} {\bibfnamefont {E.}~\bibnamefont
  {Mart\'{\i}n-Mart\'{\i}nez}}, \bibinfo {author} {\bibfnamefont
  {M.}~\bibnamefont {Montero}},\ and\ \bibinfo {author} {\bibfnamefont
  {M.}~\bibnamefont {del Rey}},\ }\bibfield  {title} {\bibinfo {title}
  {Wavepacket detection with the unruh-dewitt model},\ }\href
  {https://doi.org/10.1103/PhysRevD.87.064038} {\bibfield  {journal} {\bibinfo
  {journal} {Phys. Rev. D}\ }\textbf {\bibinfo {volume} {87}},\ \bibinfo
  {pages} {064038} (\bibinfo {year} {2013})}\BibitemShut {NoStop}%
\bibitem [{\citenamefont {Alhambra}\ \emph {et~al.}(2014)\citenamefont
  {Alhambra}, \citenamefont {Kempf},\ and\ \citenamefont
  {Mart\'{\i}n-Mart\'{\i}nez}}]{Alhambra.CasimirForces}%
  \BibitemOpen
  \bibfield  {author} {\bibinfo {author} {\bibfnamefont {A.~M.}\ \bibnamefont
  {Alhambra}}, \bibinfo {author} {\bibfnamefont {A.}~\bibnamefont {Kempf}},\
  and\ \bibinfo {author} {\bibfnamefont {E.}~\bibnamefont
  {Mart\'{\i}n-Mart\'{\i}nez}},\ }\bibfield  {title} {\bibinfo {title} {Casimir
  forces on atoms in optical cavities},\ }\href
  {https://doi.org/10.1103/PhysRevA.89.033835} {\bibfield  {journal} {\bibinfo
  {journal} {Phys. Rev. A}\ }\textbf {\bibinfo {volume} {89}},\ \bibinfo
  {pages} {033835} (\bibinfo {year} {2014})}\BibitemShut {NoStop}%
\bibitem [{\citenamefont {Pozas-Kerstjens}\ and\ \citenamefont
  {Mart\'{\i}n-Mart\'{\i}nez}(2016)}]{pozas2016entanglement}%
  \BibitemOpen
  \bibfield  {author} {\bibinfo {author} {\bibfnamefont {A.}~\bibnamefont
  {Pozas-Kerstjens}}\ and\ \bibinfo {author} {\bibfnamefont {E.}~\bibnamefont
  {Mart\'{\i}n-Mart\'{\i}nez}},\ }\bibfield  {title} {\bibinfo {title}
  {Entanglement harvesting from the electromagnetic vacuum with hydrogenlike
  atoms},\ }\href {https://doi.org/10.1103/PhysRevD.94.064074} {\bibfield
  {journal} {\bibinfo  {journal} {Phys. Rev. D}\ }\textbf {\bibinfo {volume}
  {94}},\ \bibinfo {pages} {064074} (\bibinfo {year} {2016})}\BibitemShut
  {NoStop}%
\bibitem [{\citenamefont {Valentini}(1991)}]{Valentini1991nonlocalcorr}%
  \BibitemOpen
  \bibfield  {author} {\bibinfo {author} {\bibfnamefont {A.}~\bibnamefont
  {Valentini}},\ }\bibfield  {title} {\bibinfo {title} {Non-local correlations
  in quantum electrodynamics},\ }\href
  {https://doi.org/https://doi.org/10.1016/0375-9601(91)90952-5} {\bibfield
  {journal} {\bibinfo  {journal} {Phys. Lett.}\ }\textbf {\bibinfo {volume}
  {153A}},\ \bibinfo {pages} {321 } (\bibinfo {year} {1991})}\BibitemShut
  {NoStop}%
\bibitem [{\citenamefont {Reznik}(2003)}]{reznik2003entanglement}%
  \BibitemOpen
  \bibfield  {author} {\bibinfo {author} {\bibfnamefont {B.}~\bibnamefont
  {Reznik}},\ }\bibfield  {title} {\bibinfo {title} {Entanglement from the
  vacuum},\ }\href {https://doi.org/https://doi.org/10.1023/A:1022875910744}
  {\bibfield  {journal} {\bibinfo  {journal} {Found. Phys.}\ }\textbf {\bibinfo
  {volume} {33}},\ \bibinfo {pages} {167} (\bibinfo {year} {2003})}\BibitemShut
  {NoStop}%
\bibitem [{\citenamefont {Reznik}\ \emph {et~al.}(2005)\citenamefont {Reznik},
  \citenamefont {Retzker},\ and\ \citenamefont {Silman}}]{reznik2005violating}%
  \BibitemOpen
  \bibfield  {author} {\bibinfo {author} {\bibfnamefont {B.}~\bibnamefont
  {Reznik}}, \bibinfo {author} {\bibfnamefont {A.}~\bibnamefont {Retzker}},\
  and\ \bibinfo {author} {\bibfnamefont {J.}~\bibnamefont {Silman}},\
  }\bibfield  {title} {\bibinfo {title} {{Violating Bell's inequalities in
  vacuum}},\ }\href {https://doi.org/10.1103/PhysRevA.71.042104} {\bibfield
  {journal} {\bibinfo  {journal} {Phys. Rev. A}\ }\textbf {\bibinfo {volume}
  {71}},\ \bibinfo {pages} {042104} (\bibinfo {year} {2005})}\BibitemShut
  {NoStop}%
\bibitem [{\citenamefont {Steeg}\ and\ \citenamefont
  {Menicucci}(2009)}]{Steeg2009}%
  \BibitemOpen
  \bibfield  {author} {\bibinfo {author} {\bibfnamefont {G.~V.}\ \bibnamefont
  {Steeg}}\ and\ \bibinfo {author} {\bibfnamefont {N.~C.}\ \bibnamefont
  {Menicucci}},\ }\bibfield  {title} {\bibinfo {title} {Entangling power of an
  expanding universe},\ }\href {https://doi.org/10.1103/PhysRevD.79.044027}
  {\bibfield  {journal} {\bibinfo  {journal} {Phys. Rev. D}\ }\textbf {\bibinfo
  {volume} {79}},\ \bibinfo {pages} {044027} (\bibinfo {year}
  {2009})}\BibitemShut {NoStop}%
\bibitem [{\citenamefont {Pozas-Kerstjens}\ and\ \citenamefont
  {Mart\'{\i}n-Mart\'{\i}nez}(2015)}]{pozas2015harvesting}%
  \BibitemOpen
  \bibfield  {author} {\bibinfo {author} {\bibfnamefont {A.}~\bibnamefont
  {Pozas-Kerstjens}}\ and\ \bibinfo {author} {\bibfnamefont {E.}~\bibnamefont
  {Mart\'{\i}n-Mart\'{\i}nez}},\ }\bibfield  {title} {\bibinfo {title}
  {Harvesting correlations from the quantum vacuum},\ }\href
  {https://doi.org/10.1103/PhysRevD.92.064042} {\bibfield  {journal} {\bibinfo
  {journal} {Phys. Rev. D}\ }\textbf {\bibinfo {volume} {92}},\ \bibinfo
  {pages} {064042} (\bibinfo {year} {2015})}\BibitemShut {NoStop}%
\bibitem [{\citenamefont {Maeso-Garc\'{\i}a}\ \emph {et~al.}(2022)\citenamefont
  {Maeso-Garc\'{\i}a}, \citenamefont {Polo-G\'omez},\ and\ \citenamefont
  {Mart\'{\i}n-Mart\'{\i}nez}}]{Maeso.state.covariance}%
  \BibitemOpen
  \bibfield  {author} {\bibinfo {author} {\bibfnamefont {H.}~\bibnamefont
  {Maeso-Garc\'{\i}a}}, \bibinfo {author} {\bibfnamefont {J.}~\bibnamefont
  {Polo-G\'omez}},\ and\ \bibinfo {author} {\bibfnamefont {E.}~\bibnamefont
  {Mart\'{\i}n-Mart\'{\i}nez}},\ }\bibfield  {title} {\bibinfo {title}
  {Entanglement harvesting: State dependence and covariance},\ }\href
  {https://doi.org/10.1103/PhysRevD.106.105001} {\bibfield  {journal} {\bibinfo
   {journal} {Phys. Rev. D}\ }\textbf {\bibinfo {volume} {106}},\ \bibinfo
  {pages} {105001} (\bibinfo {year} {2022})}\BibitemShut {NoStop}%
\bibitem [{\citenamefont {Mart\'{\i}n-Mart\'{\i}nez}\ \emph
  {et~al.}(2016)\citenamefont {Mart\'{\i}n-Mart\'{\i}nez}, \citenamefont
  {Smith},\ and\ \citenamefont {Terno}}]{smith2016topology}%
  \BibitemOpen
  \bibfield  {author} {\bibinfo {author} {\bibfnamefont {E.}~\bibnamefont
  {Mart\'{\i}n-Mart\'{\i}nez}}, \bibinfo {author} {\bibfnamefont {A.~R.~H.}\
  \bibnamefont {Smith}},\ and\ \bibinfo {author} {\bibfnamefont {D.~R.}\
  \bibnamefont {Terno}},\ }\bibfield  {title} {\bibinfo {title} {Spacetime
  structure and vacuum entanglement},\ }\href
  {https://doi.org/10.1103/PhysRevD.93.044001} {\bibfield  {journal} {\bibinfo
  {journal} {Phys. Rev. D}\ }\textbf {\bibinfo {volume} {93}},\ \bibinfo
  {pages} {044001} (\bibinfo {year} {2016})}\BibitemShut {NoStop}%
\bibitem [{\citenamefont {Kukita}\ and\ \citenamefont
  {Nambu}(2017)}]{kukita2017harvesting}%
  \BibitemOpen
  \bibfield  {author} {\bibinfo {author} {\bibfnamefont {S.}~\bibnamefont
  {Kukita}}\ and\ \bibinfo {author} {\bibfnamefont {Y.}~\bibnamefont {Nambu}},\
  }\bibfield  {title} {\bibinfo {title} {{Harvesting large scale entanglement
  in de Sitter space with multiple detectors}},\ }\href
  {https://doi.org/10.3390/e19090449} {\bibfield  {journal} {\bibinfo
  {journal} {Entropy}\ }\textbf {\bibinfo {volume} {19}},\ \bibinfo {pages}
  {449} (\bibinfo {year} {2017})}\BibitemShut {NoStop}%
\bibitem [{\citenamefont {Henderson}\ \emph {et~al.}(2018)\citenamefont
  {Henderson}, \citenamefont {Hennigar}, \citenamefont {Mann}, \citenamefont
  {Smith},\ and\ \citenamefont {Zhang}}]{henderson2018harvesting}%
  \BibitemOpen
  \bibfield  {author} {\bibinfo {author} {\bibfnamefont {L.~J.}\ \bibnamefont
  {Henderson}}, \bibinfo {author} {\bibfnamefont {R.~A.}\ \bibnamefont
  {Hennigar}}, \bibinfo {author} {\bibfnamefont {R.~B.}\ \bibnamefont {Mann}},
  \bibinfo {author} {\bibfnamefont {A.~R.~H.}\ \bibnamefont {Smith}},\ and\
  \bibinfo {author} {\bibfnamefont {J.}~\bibnamefont {Zhang}},\ }\bibfield
  {title} {\bibinfo {title} {Harvesting entanglement from the black hole
  vacuum},\ }\href {https://doi.org/10.1088/1361-6382/aae27e} {\bibfield
  {journal} {\bibinfo  {journal} {Classical Quantum Gravity}\ }\textbf
  {\bibinfo {volume} {35}},\ \bibinfo {pages} {21LT02} (\bibinfo {year}
  {2018})}\BibitemShut {NoStop}%
\bibitem [{\citenamefont {Ng}\ \emph {et~al.}(2018)\citenamefont {Ng},
  \citenamefont {Mann},\ and\ \citenamefont
  {Mart\'{\i}n-Mart\'{\i}nez}}]{ng2018AdS}%
  \BibitemOpen
  \bibfield  {author} {\bibinfo {author} {\bibfnamefont {K.~K.}\ \bibnamefont
  {Ng}}, \bibinfo {author} {\bibfnamefont {R.~B.}\ \bibnamefont {Mann}},\ and\
  \bibinfo {author} {\bibfnamefont {E.}~\bibnamefont
  {Mart\'{\i}n-Mart\'{\i}nez}},\ }\bibfield  {title} {\bibinfo {title}
  {{Unruh-DeWitt detectors and entanglement: The anti--de Sitter space}},\
  }\href {https://doi.org/10.1103/PhysRevD.98.125005} {\bibfield  {journal}
  {\bibinfo  {journal} {Phys. Rev. D}\ }\textbf {\bibinfo {volume} {98}},\
  \bibinfo {pages} {125005} (\bibinfo {year} {2018})}\BibitemShut {NoStop}%
\bibitem [{\citenamefont {Henderson}\ \emph {et~al.}(2019)\citenamefont
  {Henderson}, \citenamefont {Hennigar}, \citenamefont {Mann}, \citenamefont
  {Smith},\ and\ \citenamefont {Zhang}}]{henderson2019entangling}%
  \BibitemOpen
  \bibfield  {author} {\bibinfo {author} {\bibfnamefont {L.~J.}\ \bibnamefont
  {Henderson}}, \bibinfo {author} {\bibfnamefont {R.~A.}\ \bibnamefont
  {Hennigar}}, \bibinfo {author} {\bibfnamefont {R.~B.}\ \bibnamefont {Mann}},
  \bibinfo {author} {\bibfnamefont {A.~R.}\ \bibnamefont {Smith}},\ and\
  \bibinfo {author} {\bibfnamefont {J.}~\bibnamefont {Zhang}},\ }\bibfield
  {title} {\bibinfo {title} {{Entangling detectors in anti-de Sitter space}},\
  }\href {https://doi.org/https://doi.org/10.1007/JHEP05(2019)178} {\bibfield
  {journal} {\bibinfo  {journal} {J. High Energy Phys.}\ }\textbf {\bibinfo
  {volume} {05}}\bibinfo  {number} { (2019)},\ \bibinfo {pages}
  {178}}\BibitemShut {NoStop}%
\bibitem [{\citenamefont {Cong}\ \emph {et~al.}(2020)\citenamefont {Cong},
  \citenamefont {Qian}, \citenamefont {Good},\ and\ \citenamefont
  {Mann}}]{cong2020horizon}%
  \BibitemOpen
\bibfield  {number} {  }\bibfield  {author} {\bibinfo {author} {\bibfnamefont
  {W.}~\bibnamefont {Cong}}, \bibinfo {author} {\bibfnamefont {C.}~\bibnamefont
  {Qian}}, \bibinfo {author} {\bibfnamefont {M.~R.}\ \bibnamefont {Good}},\
  and\ \bibinfo {author} {\bibfnamefont {R.~B.}\ \bibnamefont {Mann}},\
  }\bibfield  {title} {\bibinfo {title} {Effects of horizons on entanglement
  harvesting},\ }\href {https://doi.org/10.1007/JHEP10(2020)067} {\bibfield
  {journal} {\bibinfo  {journal} {J. High Energy Phys.}\ }\textbf {\bibinfo
  {volume} {10}}\bibinfo  {number} { (2020)},\ \bibinfo {pages}
  {67}}\BibitemShut {NoStop}%
\bibitem [{\citenamefont {Gray}\ \emph {et~al.}(2021)\citenamefont {Gray},
  \citenamefont {Kubizňák}, \citenamefont {May}, \citenamefont {Timmerman},\
  and\ \citenamefont {Tjoa}}]{FinnShockwave}%
  \BibitemOpen
\bibfield  {number} {  }\bibfield  {author} {\bibinfo {author} {\bibfnamefont
  {F.}~\bibnamefont {Gray}}, \bibinfo {author} {\bibfnamefont {D.}~\bibnamefont
  {Kubizňák}}, \bibinfo {author} {\bibfnamefont {T.}~\bibnamefont {May}},
  \bibinfo {author} {\bibfnamefont {S.}~\bibnamefont {Timmerman}},\ and\
  \bibinfo {author} {\bibfnamefont {E.}~\bibnamefont {Tjoa}},\ }\bibfield
  {title} {\bibinfo {title} {{Quantum imprints of gravitational shockwaves}},\
  }\href {https://doi.org/https://doi.org/10.1007/JHEP11(2021)054} {\bibfield
  {journal} {\bibinfo  {journal} {J. High Energy Phys.}\ }\textbf {\bibinfo
  {volume} {11}}\bibinfo  {number} { (2021)},\ \bibinfo {pages}
  {054}}\BibitemShut {NoStop}%
\bibitem [{\citenamefont {Liu}\ \emph {et~al.}(2022)\citenamefont {Liu},
  \citenamefont {Zhang}, \citenamefont {Mann},\ and\ \citenamefont
  {Yu}}]{Liu.acceleration}%
  \BibitemOpen
\bibfield  {number} {  }\bibfield  {author} {\bibinfo {author} {\bibfnamefont
  {Z.}~\bibnamefont {Liu}}, \bibinfo {author} {\bibfnamefont {J.}~\bibnamefont
  {Zhang}}, \bibinfo {author} {\bibfnamefont {R.~B.}\ \bibnamefont {Mann}},\
  and\ \bibinfo {author} {\bibfnamefont {H.}~\bibnamefont {Yu}},\ }\bibfield
  {title} {\bibinfo {title} {Does acceleration assist entanglement
  harvesting?},\ }\href {https://doi.org/10.1103/PhysRevD.105.085012}
  {\bibfield  {journal} {\bibinfo  {journal} {Phys. Rev. D}\ }\textbf {\bibinfo
  {volume} {105}},\ \bibinfo {pages} {085012} (\bibinfo {year}
  {2022})}\BibitemShut {NoStop}%
\bibitem [{\citenamefont {Suryaatmadja}\ \emph {et~al.}(2022)\citenamefont
  {Suryaatmadja}, \citenamefont {Mann},\ and\ \citenamefont
  {Cong}}]{Diki.inertial}%
  \BibitemOpen
  \bibfield  {author} {\bibinfo {author} {\bibfnamefont {C.}~\bibnamefont
  {Suryaatmadja}}, \bibinfo {author} {\bibfnamefont {R.~B.}\ \bibnamefont
  {Mann}},\ and\ \bibinfo {author} {\bibfnamefont {W.}~\bibnamefont {Cong}},\
  }\bibfield  {title} {\bibinfo {title} {Entanglement harvesting of inertially
  moving unruh-dewitt detectors in minkowski spacetime},\ }\href
  {https://doi.org/10.1103/PhysRevD.106.076002} {\bibfield  {journal} {\bibinfo
   {journal} {Phys. Rev. D}\ }\textbf {\bibinfo {volume} {106}},\ \bibinfo
  {pages} {076002} (\bibinfo {year} {2022})}\BibitemShut {NoStop}%
\bibitem [{\citenamefont {Alsing}\ and\ \citenamefont
  {Milburn}(2003)}]{Alsing.teleportation}%
  \BibitemOpen
  \bibfield  {author} {\bibinfo {author} {\bibfnamefont {P.~M.}\ \bibnamefont
  {Alsing}}\ and\ \bibinfo {author} {\bibfnamefont {G.~J.}\ \bibnamefont
  {Milburn}},\ }\bibfield  {title} {\bibinfo {title} {Teleportation with a
  uniformly accelerated partner},\ }\href
  {https://doi.org/10.1103/PhysRevLett.91.180404} {\bibfield  {journal}
  {\bibinfo  {journal} {Phys. Rev. Lett.}\ }\textbf {\bibinfo {volume} {91}},\
  \bibinfo {pages} {180404} (\bibinfo {year} {2003})}\BibitemShut {NoStop}%
\bibitem [{\citenamefont {Fuentes-Schuller}\ and\ \citenamefont
  {Mann}(2005)}]{FuentesAliceFalls}%
  \BibitemOpen
  \bibfield  {author} {\bibinfo {author} {\bibfnamefont {I.}~\bibnamefont
  {Fuentes-Schuller}}\ and\ \bibinfo {author} {\bibfnamefont {R.~B.}\
  \bibnamefont {Mann}},\ }\bibfield  {title} {\bibinfo {title} {Alice falls
  into a black hole: Entanglement in noninertial frames},\ }\href
  {https://doi.org/10.1103/PhysRevLett.95.120404} {\bibfield  {journal}
  {\bibinfo  {journal} {Phys. Rev. Lett.}\ }\textbf {\bibinfo {volume} {95}},\
  \bibinfo {pages} {120404} (\bibinfo {year} {2005})}\BibitemShut {NoStop}%
\bibitem [{\citenamefont {Alsing}\ \emph {et~al.}(2006)\citenamefont {Alsing},
  \citenamefont {Fuentes-Schuller}, \citenamefont {Mann},\ and\ \citenamefont
  {Tessier}}]{AlsingDiracFields}%
  \BibitemOpen
  \bibfield  {author} {\bibinfo {author} {\bibfnamefont {P.~M.}\ \bibnamefont
  {Alsing}}, \bibinfo {author} {\bibfnamefont {I.}~\bibnamefont
  {Fuentes-Schuller}}, \bibinfo {author} {\bibfnamefont {R.~B.}\ \bibnamefont
  {Mann}},\ and\ \bibinfo {author} {\bibfnamefont {T.~E.}\ \bibnamefont
  {Tessier}},\ }\bibfield  {title} {\bibinfo {title} {Entanglement of dirac
  fields in noninertial frames},\ }\href
  {https://doi.org/10.1103/PhysRevA.74.032326} {\bibfield  {journal} {\bibinfo
  {journal} {Phys. Rev. A}\ }\textbf {\bibinfo {volume} {74}},\ \bibinfo
  {pages} {032326} (\bibinfo {year} {2006})}\BibitemShut {NoStop}%
\bibitem [{\citenamefont {Lin}\ \emph {et~al.}(2008)\citenamefont {Lin},
  \citenamefont {Chou},\ and\ \citenamefont {Hu}}]{Lin.Disentanglement2008}%
  \BibitemOpen
  \bibfield  {author} {\bibinfo {author} {\bibfnamefont {S.-Y.}\ \bibnamefont
  {Lin}}, \bibinfo {author} {\bibfnamefont {C.-H.}\ \bibnamefont {Chou}},\ and\
  \bibinfo {author} {\bibfnamefont {B.~L.}\ \bibnamefont {Hu}},\ }\bibfield
  {title} {\bibinfo {title} {Disentanglement of two harmonic oscillators in
  relativistic motion},\ }\href {https://doi.org/10.1103/PhysRevD.78.125025}
  {\bibfield  {journal} {\bibinfo  {journal} {Phys. Rev. D}\ }\textbf {\bibinfo
  {volume} {78}},\ \bibinfo {pages} {125025} (\bibinfo {year}
  {2008})}\BibitemShut {NoStop}%
\bibitem [{\citenamefont {Lin}\ and\ \citenamefont
  {Hu}(2009)}]{Lin.Temporal.2009}%
  \BibitemOpen
  \bibfield  {author} {\bibinfo {author} {\bibfnamefont {S.-Y.}\ \bibnamefont
  {Lin}}\ and\ \bibinfo {author} {\bibfnamefont {B.~L.}\ \bibnamefont {Hu}},\
  }\bibfield  {title} {\bibinfo {title} {Temporal and spatial dependence of
  quantum entanglement from a field theory perspective},\ }\href
  {https://doi.org/10.1103/PhysRevD.79.085020} {\bibfield  {journal} {\bibinfo
  {journal} {Phys. Rev. D}\ }\textbf {\bibinfo {volume} {79}},\ \bibinfo
  {pages} {085020} (\bibinfo {year} {2009})}\BibitemShut {NoStop}%
\bibitem [{\citenamefont {Landulfo}\ and\ \citenamefont
  {Matsas}(2009)}]{Landulfo2009suddendeath}%
  \BibitemOpen
  \bibfield  {author} {\bibinfo {author} {\bibfnamefont {A.~G.~S.}\
  \bibnamefont {Landulfo}}\ and\ \bibinfo {author} {\bibfnamefont {G.~E.~A.}\
  \bibnamefont {Matsas}},\ }\bibfield  {title} {\bibinfo {title} {{Sudden death
  of entanglement and teleportation fidelity loss via the Unruh effect}},\
  }\href {https://doi.org/10.1103/PhysRevA.80.032315} {\bibfield  {journal}
  {\bibinfo  {journal} {Phys. Rev. A}\ }\textbf {\bibinfo {volume} {80}},\
  \bibinfo {pages} {032315} (\bibinfo {year} {2009})}\BibitemShut {NoStop}%
\bibitem [{\citenamefont {Doukas}\ and\ \citenamefont
  {Carson}(2010)}]{Doukas.orbit.PhysRevA.81.062320}%
  \BibitemOpen
  \bibfield  {author} {\bibinfo {author} {\bibfnamefont {J.}~\bibnamefont
  {Doukas}}\ and\ \bibinfo {author} {\bibfnamefont {B.}~\bibnamefont
  {Carson}},\ }\bibfield  {title} {\bibinfo {title} {Entanglement of two qubits
  in a relativistic orbit},\ }\href
  {https://doi.org/10.1103/PhysRevA.81.062320} {\bibfield  {journal} {\bibinfo
  {journal} {Phys. Rev. A}\ }\textbf {\bibinfo {volume} {81}},\ \bibinfo
  {pages} {062320} (\bibinfo {year} {2010})}\BibitemShut {NoStop}%
\bibitem [{\citenamefont {C\'eleri}\ \emph {et~al.}(2010)\citenamefont
  {C\'eleri}, \citenamefont {Landulfo}, \citenamefont {Serra},\ and\
  \citenamefont {Matsas}}]{Landulfo.SuddenChange}%
  \BibitemOpen
  \bibfield  {author} {\bibinfo {author} {\bibfnamefont {L.~C.}\ \bibnamefont
  {C\'eleri}}, \bibinfo {author} {\bibfnamefont {A.~G.~S.}\ \bibnamefont
  {Landulfo}}, \bibinfo {author} {\bibfnamefont {R.~M.}\ \bibnamefont
  {Serra}},\ and\ \bibinfo {author} {\bibfnamefont {G.~E.~A.}\ \bibnamefont
  {Matsas}},\ }\bibfield  {title} {\bibinfo {title} {Sudden change in quantum
  and classical correlations and the unruh effect},\ }\href
  {https://doi.org/10.1103/PhysRevA.81.062130} {\bibfield  {journal} {\bibinfo
  {journal} {Phys. Rev. A}\ }\textbf {\bibinfo {volume} {81}},\ \bibinfo
  {pages} {062130} (\bibinfo {year} {2010})}\BibitemShut {NoStop}%
\bibitem [{\citenamefont {Ostapchuk}\ \emph {et~al.}(2012)\citenamefont
  {Ostapchuk}, \citenamefont {Lin}, \citenamefont {Mann},\ and\ \citenamefont
  {Hu}}]{Ostapchuk.entanglement.dynamics}%
  \BibitemOpen
  \bibfield  {author} {\bibinfo {author} {\bibfnamefont {D.~C.~M.}\
  \bibnamefont {Ostapchuk}}, \bibinfo {author} {\bibfnamefont {S.-Y.}\
  \bibnamefont {Lin}}, \bibinfo {author} {\bibfnamefont {R.~B.}\ \bibnamefont
  {Mann}},\ and\ \bibinfo {author} {\bibfnamefont {B.~L.}\ \bibnamefont {Hu}},\
  }\bibfield  {title} {\bibinfo {title} {{Entanglement dynamics between
  inertial and non-uniformly accelerated detectors}},\ }\href
  {https://doi.org/https://doi.org/10.1007/JHEP07(2012)072} {\bibfield
  {journal} {\bibinfo  {journal} {J. High Energy Phys.}\ }\textbf {\bibinfo
  {volume} {07}}\bibinfo  {number} { (2012)},\ \bibinfo {pages}
  {072}}\BibitemShut {NoStop}%
\bibitem [{\citenamefont {Mart\'{\i}n-Mart\'{\i}nez}\ and\ \citenamefont
  {Louko}(2015)}]{EMM.Jorma.Firewall}%
  \BibitemOpen
\bibfield  {number} {  }\bibfield  {author} {\bibinfo {author} {\bibfnamefont
  {E.}~\bibnamefont {Mart\'{\i}n-Mart\'{\i}nez}}\ and\ \bibinfo {author}
  {\bibfnamefont {J.}~\bibnamefont {Louko}},\ }\bibfield  {title} {\bibinfo
  {title} {$(1+1)\mathrm{D}$ calculation provides evidence that quantum
  entanglement survives a firewall},\ }\href
  {https://doi.org/10.1103/PhysRevLett.115.031301} {\bibfield  {journal}
  {\bibinfo  {journal} {Phys. Rev. Lett.}\ }\textbf {\bibinfo {volume} {115}},\
  \bibinfo {pages} {031301} (\bibinfo {year} {2015})}\BibitemShut {NoStop}%
\bibitem [{\citenamefont {Rodríguez-Camargo}\ \emph
  {et~al.}(2018)\citenamefont {Rodríguez-Camargo}, \citenamefont {Menezes},\
  and\ \citenamefont {Svaiter}}]{Rodriguez.finite.time}%
  \BibitemOpen
  \bibfield  {author} {\bibinfo {author} {\bibfnamefont {C.}~\bibnamefont
  {Rodríguez-Camargo}}, \bibinfo {author} {\bibfnamefont {G.}~\bibnamefont
  {Menezes}},\ and\ \bibinfo {author} {\bibfnamefont {N.}~\bibnamefont
  {Svaiter}},\ }\bibfield  {title} {\bibinfo {title} {Finite-time response
  function of uniformly accelerated entangled atoms},\ }\href
  {https://doi.org/https://doi.org/10.1016/j.aop.2018.07.002} {\bibfield
  {journal} {\bibinfo  {journal} {Annals of Physics}\ }\textbf {\bibinfo
  {volume} {396}},\ \bibinfo {pages} {266} (\bibinfo {year}
  {2018})}\BibitemShut {NoStop}%
\bibitem [{\citenamefont {Gallock-Yoshimura}\ and\ \citenamefont
  {Mann}(2021)}]{GallockEntangledDetectors}%
  \BibitemOpen
  \bibfield  {author} {\bibinfo {author} {\bibfnamefont {K.}~\bibnamefont
  {Gallock-Yoshimura}}\ and\ \bibinfo {author} {\bibfnamefont {R.~B.}\
  \bibnamefont {Mann}},\ }\bibfield  {title} {\bibinfo {title} {Entangled
  detectors nonperturbatively harvest mutual information},\ }\href
  {https://doi.org/10.1103/PhysRevD.104.125017} {\bibfield  {journal} {\bibinfo
   {journal} {Phys. Rev. D}\ }\textbf {\bibinfo {volume} {104}},\ \bibinfo
  {pages} {125017} (\bibinfo {year} {2021})}\BibitemShut {NoStop}%
\bibitem [{\citenamefont {Chowdhury}\ and\ \citenamefont
  {Majhi}(2022)}]{Chowdhury.FateEntanglement}%
  \BibitemOpen
  \bibfield  {author} {\bibinfo {author} {\bibfnamefont {P.}~\bibnamefont
  {Chowdhury}}\ and\ \bibinfo {author} {\bibfnamefont {B.~R.}\ \bibnamefont
  {Majhi}},\ }\bibfield  {title} {\bibinfo {title} {{Fate of entanglement
  between two Unruh-DeWitt detectors due to their motion and background
  temperature}},\ }\href
  {https://doi.org/https://doi.org/10.1007/JHEP05(2022)025} {\bibfield
  {journal} {\bibinfo  {journal} {J. High Energy Phys.}\ }\textbf {\bibinfo
  {volume} {05}}\bibinfo  {number} { (2022)},\ \bibinfo {pages}
  {025}}\BibitemShut {NoStop}%
\bibitem [{\citenamefont {Soares}\ \emph {et~al.}(2022)\citenamefont {Soares},
  \citenamefont {Menezes},\ and\ \citenamefont
  {Svaiter}}]{Soares.Entanglement.dynamics}%
  \BibitemOpen
\bibfield  {number} {  }\bibfield  {author} {\bibinfo {author} {\bibfnamefont
  {M.~S.}\ \bibnamefont {Soares}}, \bibinfo {author} {\bibfnamefont
  {G.}~\bibnamefont {Menezes}},\ and\ \bibinfo {author} {\bibfnamefont {N.~F.}\
  \bibnamefont {Svaiter}},\ }\href {https://doi.org/10.48550/ARXIV.2205.11628}
  {\bibinfo {title} {Entanglement dynamics: Generalized master equation for
  uniformly accelerated two-level systems}} (\bibinfo {year}
  {2022})\BibitemShut {NoStop}%
\bibitem [{\citenamefont {Tjoa}\ and\ \citenamefont
  {Mart\'{\i}n-Mart\'{\i}nez}(2021)}]{TjoaSignal}%
  \BibitemOpen
  \bibfield  {author} {\bibinfo {author} {\bibfnamefont {E.}~\bibnamefont
  {Tjoa}}\ and\ \bibinfo {author} {\bibfnamefont {E.}~\bibnamefont
  {Mart\'{\i}n-Mart\'{\i}nez}},\ }\bibfield  {title} {\bibinfo {title} {When
  entanglement harvesting is not really harvesting},\ }\href
  {https://doi.org/10.1103/PhysRevD.104.125005} {\bibfield  {journal} {\bibinfo
   {journal} {Phys. Rev. D}\ }\textbf {\bibinfo {volume} {104}},\ \bibinfo
  {pages} {125005} (\bibinfo {year} {2021})}\BibitemShut {NoStop}%
\bibitem [{\citenamefont {Cliche}\ and\ \citenamefont
  {Kempf}(2011)}]{Cliche.weak.gravity}%
  \BibitemOpen
  \bibfield  {author} {\bibinfo {author} {\bibfnamefont {M.}~\bibnamefont
  {Cliche}}\ and\ \bibinfo {author} {\bibfnamefont {A.}~\bibnamefont {Kempf}},\
  }\bibfield  {title} {\bibinfo {title} {Vacuum entanglement enhancement by a
  weak gravitational field},\ }\href
  {https://doi.org/10.1103/PhysRevD.83.045019} {\bibfield  {journal} {\bibinfo
  {journal} {Phys. Rev. D}\ }\textbf {\bibinfo {volume} {83}},\ \bibinfo
  {pages} {045019} (\bibinfo {year} {2011})}\BibitemShut {NoStop}%
\bibitem [{\citenamefont {Dray}\ and\ \citenamefont {{'t
  Hooft}}(1985)}]{Dray.shockwave}%
  \BibitemOpen
  \bibfield  {author} {\bibinfo {author} {\bibfnamefont {T.}~\bibnamefont
  {Dray}}\ and\ \bibinfo {author} {\bibfnamefont {G.}~\bibnamefont {{'t
  Hooft}}},\ }\bibfield  {title} {\bibinfo {title} {The gravitational shock
  wave of a massless particle},\ }\href
  {https://doi.org/https://doi.org/10.1016/0550-3213(85)90525-5} {\bibfield
  {journal} {\bibinfo  {journal} {Nuclear Physics B}\ }\textbf {\bibinfo
  {volume} {253}},\ \bibinfo {pages} {173} (\bibinfo {year}
  {1985})}\BibitemShut {NoStop}%
\bibitem [{\citenamefont {Aichelburg}\ and\ \citenamefont
  {Sexl}(1971)}]{Aichelburg1971gravitational}%
  \BibitemOpen
  \bibfield  {author} {\bibinfo {author} {\bibfnamefont {P.~C.}\ \bibnamefont
  {Aichelburg}}\ and\ \bibinfo {author} {\bibfnamefont {R.~U.}\ \bibnamefont
  {Sexl}},\ }\bibfield  {title} {\bibinfo {title} {On the gravitational field
  of a massless particle},\ }\href
  {https://link.springer.com/article/10.1007/BF00758149} {\bibfield  {journal}
  {\bibinfo  {journal} {General Relativity and Gravitation}\ }\textbf {\bibinfo
  {volume} {2}},\ \bibinfo {pages} {303} (\bibinfo {year} {1971})}\BibitemShut
  {NoStop}%
\bibitem [{\citenamefont {Brinkmann}(1923)}]{Brinkmann1923riemann}%
  \BibitemOpen
  \bibfield  {author} {\bibinfo {author} {\bibfnamefont {H.}~\bibnamefont
  {Brinkmann}},\ }\bibfield  {title} {\bibinfo {title} {On riemann spaces
  conformal to euclidean space},\ }\href {https://doi.org/10.1073/pnas.9.1.1}
  {\bibfield  {journal} {\bibinfo  {journal} {Proceedings of the National
  Academy of Sciences of the United States of America}\ ,\ \bibinfo {pages}
  {1}} (\bibinfo {year} {1923})}\BibitemShut {NoStop}%
\bibitem [{\citenamefont {Loustó}\ and\ \citenamefont
  {Sánchez}(1991)}]{Lousto.domainwall}%
  \BibitemOpen
  \bibfield  {author} {\bibinfo {author} {\bibfnamefont {C.}~\bibnamefont
  {Loustó}}\ and\ \bibinfo {author} {\bibfnamefont {N.}~\bibnamefont
  {Sánchez}},\ }\bibfield  {title} {\bibinfo {title} {Gravitational shock
  waves generated by extended sources: ultrarelativistic cosmic strings,
  monopoles and domain walls},\ }\href
  {https://doi.org/https://doi.org/10.1016/0550-3213(91)90311-K} {\bibfield
  {journal} {\bibinfo  {journal} {Nuclear Physics B}\ }\textbf {\bibinfo
  {volume} {355}},\ \bibinfo {pages} {231} (\bibinfo {year}
  {1991})}\BibitemShut {NoStop}%
\bibitem [{\citenamefont {Simidzija}\ and\ \citenamefont
  {Mart\'{\i}n-Mart\'{\i}nez}(2017)}]{Simidzija.Nonperturbative}%
  \BibitemOpen
  \bibfield  {author} {\bibinfo {author} {\bibfnamefont {P.}~\bibnamefont
  {Simidzija}}\ and\ \bibinfo {author} {\bibfnamefont {E.}~\bibnamefont
  {Mart\'{\i}n-Mart\'{\i}nez}},\ }\bibfield  {title} {\bibinfo {title}
  {Nonperturbative analysis of entanglement harvesting from coherent field
  states},\ }\href {https://doi.org/10.1103/PhysRevD.96.065008} {\bibfield
  {journal} {\bibinfo  {journal} {Phys. Rev. D}\ }\textbf {\bibinfo {volume}
  {96}},\ \bibinfo {pages} {065008} (\bibinfo {year} {2017})}\BibitemShut
  {NoStop}%
\bibitem [{\citenamefont {Simidzija}\ \emph {et~al.}(2018)\citenamefont
  {Simidzija}, \citenamefont {Jonsson},\ and\ \citenamefont
  {Mart\'{\i}n-Mart\'{\i}nez}}]{Simidzija2018no-go}%
  \BibitemOpen
  \bibfield  {author} {\bibinfo {author} {\bibfnamefont {P.}~\bibnamefont
  {Simidzija}}, \bibinfo {author} {\bibfnamefont {R.~H.}\ \bibnamefont
  {Jonsson}},\ and\ \bibinfo {author} {\bibfnamefont {E.}~\bibnamefont
  {Mart\'{\i}n-Mart\'{\i}nez}},\ }\bibfield  {title} {\bibinfo {title} {General
  no-go theorem for entanglement extraction},\ }\href
  {https://doi.org/10.1103/PhysRevD.97.125002} {\bibfield  {journal} {\bibinfo
  {journal} {Phys. Rev. D}\ }\textbf {\bibinfo {volume} {97}},\ \bibinfo
  {pages} {125002} (\bibinfo {year} {2018})}\BibitemShut {NoStop}%
\bibitem [{\citenamefont {Barman}\ \emph {et~al.}(2022)\citenamefont {Barman},
  \citenamefont {Choudhury}, \citenamefont {Kad},\ and\ \citenamefont
  {Majhi}}]{Barman.spontaneous}%
  \BibitemOpen
  \bibfield  {author} {\bibinfo {author} {\bibfnamefont {D.}~\bibnamefont
  {Barman}}, \bibinfo {author} {\bibfnamefont {A.}~\bibnamefont {Choudhury}},
  \bibinfo {author} {\bibfnamefont {B.}~\bibnamefont {Kad}},\ and\ \bibinfo
  {author} {\bibfnamefont {B.~R.}\ \bibnamefont {Majhi}},\ }\href
  {https://doi.org/10.48550/ARXIV.2211.00383} {\bibinfo {title} {Spontaneous
  entanglement leakage of two static entangled unruh-dewitt detectors}}
  (\bibinfo {year} {2022})\BibitemShut {NoStop}%
\end{thebibliography}%

\end{document}